\documentclass[journal]{IEEEtran} 
\ifCLASSINFOpdf
  \usepackage[pdftex]{graphicx}
\else
  \usepackage[dvips]{graphicx}
\fi
\usepackage[cmex10]{amsmath}
\usepackage[caption=false,font=footnotesize,justification=centerlast,singlelinecheck=false]{subfig}
\usepackage{fixltx2e}

\usepackage{epstopdf}
\usepackage{booktabs}
\usepackage{dcolumn}
\usepackage{amsfonts}
\usepackage{amssymb}
\usepackage{algorithm}
\usepackage{algpseudocode}
\usepackage{pstricks}

\def\x{{\mathbf x}}

\def\y{{\mathbf y}}
\def\w{{\mathbf w}}
\def\m{{\mathbf m}}
\def\n{{\mathbf n}}
\def\v{{\mathbf v}}
\def\p{{\mathbf p}}
\def\s{{\mathbf s}}
\def\d{{\mathbf d}}
\def\xe{{\mathbf x_{\mathbf e}}}
\def\xr{{\mathbf x_{\mathbf r}}}
\def\M{{\mathbf M}}
\def\H{{\mathbf H}}
\def\F{{\mathbf F}}
\def\C{{\mathbf C}}
\def\R{{\mathbf R}}
\def\Dt{{\mathbf D_{\mathbf t}}}
\def\I{{\mathbf I}}
\def\Kt{{\mathbf K_{\mathbf t}(\v)}}
\def\Kts{{\mathbf K_{\mathbf t}(\v^*)}}
\def\K{{\mathbf K}}

\def\Ps{{\mathbf \Psi}}
\DeclareMathOperator*{\argmin}{arg\,min}
\DeclareMathOperator{\TV}{TV}
\DeclareMathOperator{\soft}{\mathfrak{S}}
\DeclareMathOperator{\upd}{Update\ }

\begin{document}
%
\title{Compressed Sensing for Moving Imagery in Medical Imaging}

\author{
\IEEEauthorblockN{\c{C}a\u{g}da\c{s}~Bilen, Yao~Wang and Ivan~Selesnick}
\\
\IEEEauthorblockA{cbilen@vision.poly.edu, yao@poly.edu, selesi@poly.edu\\
Department of Electrical and Computer Engineering, 
Polytechnic Institute of NYU\\ 
5 Metrotech Center, Brooklyn, NY, 11201, USA, Telephone: (718) 260-3590}
}


%
%


\maketitle

\begin{abstract}
Numerous applications in signal processing have benefited from the theory of compressed sensing which shows that it is possible to reconstruct signals sampled below the Nyquist rate when certain conditions are satisfied. One of these conditions is that there exists a known transform that represents the signal with a sufficiently small number of non-zero coefficients. However when the signal to be reconstructed is composed of moving images or volumes, it is challenging to form such regularization constraints with traditional transforms such as wavelets. In this paper, we present a motion compensating prior for such signals that is derived directly from the optical flow constraint and can utilize the motion information during compressed sensing reconstruction. Proposed regularization method can be used in a wide variety of applications involving compressed sensing and images or volumes of moving and deforming objects. It is also shown that it is possible to estimate the signal and the motion jointly or separately. Practical examples from magnetic resonance imaging has been presented to demonstrate the benefit of the proposed method.
\end{abstract}

%
\IEEEpeerreviewmaketitle

\section{Introduction}
System inversion problems in imaging has been extensively studied for decades and the developments in compressed sensing approach in the last decade has provided a new perspective and opportunities for solution of these problems. By making use of sparse reconstruction principles, it is shown that compressed sensing enables reliable recovery of signals even if system response is measured at a rate below the Nyquist rate \cite{Candes2008}. Numerous fields has benefited from this discovery especially if data acquisition is limited due to constraints.

Medical imaging is one of the application areas that adopted compressed sensing principles rather quickly. It is shown that in magnetic resonance imaging (MRI), the number of measurement samples and thus the scan time can be reduced while preserving image quality if compressed sensing principles are used \cite{Lustig2008}. This is further improved by parallel imaging with algorithms such as SparseSENSE utilizing multiple coils \cite{Liu2008}. Although acquisition of compressed measurements are not as straightforward as MRI in X-ray computer tomography (CT), compressed sensing is shown to help for partial reconstruction from a subset of measurements \cite{Yu2009}. Photo-acoustic imaging is another application for which compressed sensing improves reconstructed image quality \cite{Provost2009}. These initial studies with compressed sensing were on still images or volumes and spatial total variation (TV) or wavelets were used as regularization constraints. 

Dynamic imaging is shown to benefit from compressed sensing even further due to the images being significantly correlated along temporal dimension and therefore represented with sparse temporal transforms. Similarly to the spatial case, temporal TV and wavelets are commonly used in dynamic CT and MRI \cite{Chen2009}, \cite{Gamper2008}. The temporal Fourier transform is also utilized with k-t SparseSENSE \cite{Otazo2010} especially with cardiac MRI, due to the cardiac motion being periodic therefore sparse with respect to Fourier transform. K-t Group Sparse SENSE is introduced by Usman et al. \cite{Usman2011} which groups pixels with respect to their spatio temporal activity to treat static and dynamic regions differently during reconstruction. Alternatively Kalman filter is utilized by Vaswani \cite{Vaswani2008} providing quick sequential reconstruction of images in time. However when there is motion in the observed object such as in cardiac imaging or due to patient motion during the acquisition, these temporal regularization methods do not always sufficiently sparsify the signal. Therefore methods compensating for motion have been investigated.

The compression of images or volumes of moving objects in time is a well-investigated field and many efficient video compression algorithms exist today such as H.264/AVC to exploit temporal dependence in video. The common approach in video compression is sequentially estimating each frame from available reference frames by estimating the motion in between. Considering the high compression achieved by video compression methods, there are algorithms that approach the compressed sensing reconstruction problem similarly. Some basic methods using information from an existing reference frame during reconstruction are presented in \cite{Lam2010}. This is further improved by Jung et al. utilizing multiple reference frames and motion compensation in k-t FOCUSS \cite{Jung2009a}, \cite{Jung2010} or other works such as \cite{Chen2011}. One of the drawbacks of these reference frame based reconstructions is that, unlike the video compression framework, having good quality reference frames in compressed sensing is not realistic in practice where the entire signal is reconstructed simultaneously not sequentially. In this paper we propose a motion compensating prior that significantly sparsifies multidimensional signals that involves motion therefore enabling high quality compressed sensing reconstruction. The proposed prior is a natural extension of the optical flow constraint and can be considered as a generalization of temporal TV. The motion is estimated either separately from or jointly with the signal. The proposed prior can be used with a variety of compressed sensing applications that involve moving objects.

The paper is organized as follows; the formulation of compressed sensing reconstruction and some recent approaches are summarized in Section~\ref{sec:ProbDef}, the proposed reconstruction method is presented in Section~\ref{sec:PropMethod}, some experimental results with dynamic MRI are reported in Section~\ref{sec:Results} and finally the conclusions and future work are discussed in Section~\ref{sec:Conc}.

\section{Compressed Sensing Reconstruction}
\label{sec:ProbDef}
Compressed sensing can be formulated as a system inversion problem in general with
\begin{equation}
\y = \H \x + \n
\end{equation}
where $\y$ is the observation or measurement vector, $\x$ is the signal to be reconstructed, $\n$ is the additive noise and $\H$ is a matrix defining the linear operations on $\x$. $\H$ can take on different forms depending on the application at hand but it is assumed to be ill conditioned and therefore there are infinitely many solutions for $\x$. Compressed sensing shows that the signal $\x$ can be recovered almost certainly if \cite{Candes2008}, \cite{Lustig2008},
\begin{enumerate}
\item $\Ps\x$ is sufficiently sparse for a known transform $\Ps$ ($\Ps'\Ps=\Ps\Ps'=\I$, $.'$ indicating the conjugate transpose)
\item $\H$ and $\Ps$ are sufficiently incoherent, i.e. the off diagonal entries of $\Ps\H'\H\Ps'$ after normalization are sufficiently small
\end{enumerate}
One of the breakthroughs in compressed sensing is that the 2nd condition is shown to be satisfied when $\H$ is formed by random entries, or by a random subset of rows of a full rank matrix regardless of $\Ps$. The 1st condition is also relaxed such that $\Ps$ can be an overcomplete transform or dictionary or a penalty operator as in case of total variation \cite{Rauhut2008}, \cite{Candes2010}. Under these conditions it is shown that $\x$ can be recovered almost certainly with the minimization 
\begin{align}
\label{eqn:L0}
\tilde{\w} &= \argmin_{\w} \Vert\w\Vert_0 \text{   s.t.  } \Vert \y - \H\Ps'\w \Vert^2_2 \leq \epsilon \\
\tilde{\x} &= \Ps'\tilde{\w}
\end{align}
in which $\Vert.\Vert_0$ is the $L_0$-norm (which is in fact a quasi-norm) that is the number of non-zero entries of a vector ($\Vert \x \Vert_p \triangleq \left(\sum\limits_i x_i^p\right)^{1/p} $ for $p>0$). The constraint $\Vert \y - \H\Ps'\w \Vert^2_2 \leq  \epsilon$ ensures consistency with the measurements and it is optimum for i.i.d. Gaussian noise with standard deviation $\epsilon$, although it has been shown that it works fairly well for other noise distributions such as Poisson. The minimization in  (\ref{eqn:L0}) is non-convex and therefore practical methods do not guarantee convergence to global minimum. It is shown in the sparse reconstruction literature that the minimization of $L_1$-norm as in
\begin{align}
\label{eqn:L1}
\tilde{\w} &= \argmin_{\w} \Vert\w\Vert_1 \text{   s.t.  } \Vert \y - \H\Ps'\w \Vert^2_2 \leq  \epsilon \\
\tilde{\x} &= \Ps'\tilde{\w}
\end{align}
will lead to the same solution as in (\ref{eqn:L0}) provided that $\w$ is sufficiently sparse \cite{Donoho2009}. The implications of this equivalence is significant since (\ref{eqn:L1}) is a convex optimization problem with a global minimum and can be solved very efficiently with methods such as Alternating Direction Method of Multipliers (ADMM) \cite{Boyd2011}, \cite{Goldstein2009} or C-SALSA \cite{Afonso2011}. In order to further simplify the minimization, the constraint in  (\ref{eqn:L1}) can be removed to form the unconstrained formulation
\begin{align}
\label{eqn:L1unconst}
\tilde{\w} &= \argmin_{\w} \lambda \Vert\w\Vert_1 + \dfrac{1}{2}\Vert \y - \H\Ps'\w \Vert^2_2 \\
\tilde{\x} &= \Ps'\tilde{\w}
\end{align}
A number of fast minimization algorithms exist for the solution of unconstrained minimization in (\ref{eqn:L1unconst}) such as TwIST \cite{Bioucas2007}, FISTA \cite{Beck2009}, ADMM and SALSA \cite{Afonso2010a} and the result is equivalent to (\ref{eqn:L1}) if $\lambda$ is adjusted properly. 

The synthesis prior formulation in (\ref{eqn:L1}) and (\ref{eqn:L1unconst}) is used with overcomplete transforms such as wavelets or non-orthogonal dictionaries. An alternative approach uses an analysis prior formulation with a penalty term for regularization as in
\begin{equation}
\label{eqn:L1an_pri}
\tilde{\x} = \argmin_{\x} \lambda R(\x) + \dfrac{1}{2}\Vert \y - \H\x \Vert^2_2
\end{equation}
in which $R(\x)$ is the penalty function that is large when $\x$ has characteristics different from a prior knowledge of $\x$. A commonly used example for such penalty functions is the total variation (TV) which is defined for 1D signals as
\begin{equation}
\TV(\x) \triangleq \sum\limits_i |x_i - x_{i-1}|
\end{equation}
and penalizes the signal gradient. Let us define our signal as concatenation of vectors of consecutive images (or volumes), $\x_i$,  such that
\begin{equation}
\x = [\x_1^T \cdots \x_{n_t}^T]^T
\end{equation}
where $.^T$ represents non-conjugate transpose and $n_t$ is the number of frames. Therefore the total variation in temporal dimension penalizing the temporal gradient can be defined as
\begin{align}
\label{eqn:TV}
\nonumber
\TV_t(\x) &= \sum_{i=2}^{n_t} \Vert\x_i - \x_{i-1}\Vert_1 \\
\nonumber &=  
\left\Vert
\begin{bmatrix} 
-\I & \I & 0 & \cdots & 0 \\
0 & -\I & \I & \cdots & 0 \\
\vdots & & \ddots & & \vdots \\
0 & \cdots & 0 & -\I & \I
\end{bmatrix}
\left[ \begin{array}{c} \x_1 \\ \vdots \\ \x_{n_t} \end{array} \right] 
\right\Vert_1 \\
&= \Vert\Dt\x\Vert_1
\end{align} 
in which $\Dt$ is the temporal difference matrix. Minimization of (\ref{eqn:L1an_pri}) with $R(\x)=\TV_t(\x)$ is possible with algorithms such as MFISTA \cite{Beck2009} or ADMM, and can be sufficient for signals with insignificant motion or large static areas between consecutive frames. However when signals have substantial motion, the frame difference may not be sufficiently sparse. 

In order to improve the sparsity of regularization and therefore the accuracy of reconstruction when dealing with signals with motion, alternative reconstruction methods that make use of motion compensation have been proposed \cite{Lam2010}, \cite{Jung2009a}, \cite{Jung2010}. The common approach in these methods is using motion compensation to create an initial estimate for the signal from available data and then using this estimate as a prior during the reconstruction. A simple example to this approach is estimating a number of frames between two reference frames by motion interpolation assuming the motion in between is linear \cite{Jung2010}. The estimate for these frames $\xe=[\xe_1^T \cdots \xe_{n_t}^T]^T$ can then be used to create a more sparse error signal $\xr = \x-\xe$, and the entire signal can be reconstructed by the minimization
\begin{align}
\label{eqn:Residual}
\tilde{\xr} &= \argmin_{\xr} \lambda R(\xr)+\dfrac{1}{2}\Vert \tilde{\y} - \H\xr\Vert^2_2 \\
\nonumber & \text{   where   } \tilde{\y} \triangleq \y - \H\xe\\
\tilde{\x} &= \xe + \tilde{\xr}
\end{align} 
in which $\xe$ is constant. $R(.)$ is either just the $L_1$-norm (assuming $\xr$ is sufficiently sparse) or a penalty function such as TV. Various methods to estimate $\xe$ have been proposed in \cite{Lam2010}, \cite{Jung2009a}, \cite{Jung2010}, most of which makes use of reference frames or low resolution reconstruction of the frames. However this approach may be inefficient for mainly two reasons:
\begin{enumerate}
\item \textit{Inaccurate $\xe$:} Unlike video compression, accurate reference frames are often not present in compressed sensing and therefore $\xe$ may not be accurate enough to result in a $\xr$ with sufficient sparsity. 
\item \textit{Signal-to-noise ratio (SNR) penalty:} Although (\ref{eqn:Residual}) is very similar to (\ref{eqn:L1an_pri}), it can be seen that $\y$ is replaced with $\tilde{\y} = \y - \H\xe$ which may significantly affect the result unless sparsity is vastly increased to compensate considering the SNR of the measurements is reduced ($E(\y)/E(\n) >> E(\tilde{\y})/E(\n)$) \cite{Akcakaya2010}.
\end{enumerate}
Therefore a better utilization of motion during reconstruction is preferable in order to avoid these drawbacks.

\section{A Motion Compensating Prior - \textit{Motion-TV} }
\label{sec:PropMethod}

\begin{figure}
\centering
\psscalebox{0.7}{
\psset{unit=0.3cm}
\input{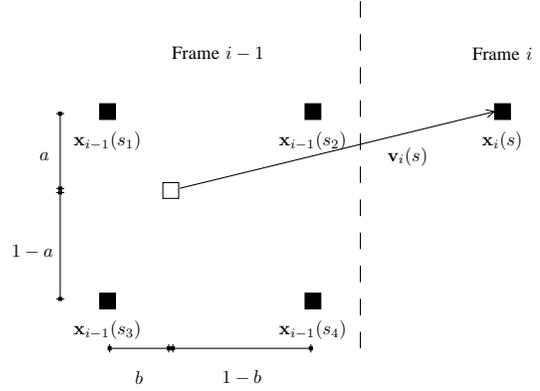}
}
\caption{The pixel $\x_i(s)$ and the motion vector $\v_i$ associated with it in Frame $i$}
\label{fig:K_example}
\end{figure}

Let us assume that the motion vectors between the frames of a signal are known but not the signal itself. Without the reference frames the motion information would be useless for reconstruction in video compression whereas in compressed sensing this information could be quite helpful. One of the main assumptions for motion estimation between images is the optical flow constraint (OFC) 
\begin{equation}
\label{eqn:OFC}
x_i(s) = x_{i-1}(s+v_i(s))
\end{equation}
with $s$ being the spatial index, which basically states that for a spatially continuous signal the signal intensity is constant along the motion path in consecutive frames. In case of discrete signals this equation does not exactly hold due to interpolation error and therefore becomes
\begin{equation}
\label{eqn:OFC_discrete}
\x_i \cong  \K_{i,i-1}\x_{i-1} 
\end{equation}
where $\K_{i,i-1}=\K(\v_i)$ is a matrix that represents the geometric transform and interpolation defined by the motion vectors , $\v_i$, between frames $i$ and $(i-1)$ and therefore $\K_{i,i-1}\x_{i-1}$ is the motion compensated estimate of $\x_i$ from $\x_{i-1}$. If all motion vectors correspond to integer pixel locations, $\K_{i,i-1}$ is simply a matrix with each row having a single non-zero entry equal to 1. In the case when motion vectors point to non-integer locations, rows of $\K_{i,i-1}$ can have multiple non zero entries depending on the interpolation method (a maximum of four non-zero entries for bilinear interpolation, 16 non-zero entries for bicubic interpolation, etc.) representing a weighted average of nearby pixel locations. An example of $\K_{i,i-1}$ can be seen in
\begin{equation}
\begin{tiny}
\label{eqn:K_example}
\underbrace{
\begin{bmatrix}
\vdots \\
\x_i(s) \\
\vdots 
\end{bmatrix}
}_{\x_i} 
\cong 
\underbrace{
\begin{bmatrix}
  &        &  & \cdots & &  &        & \\
0 & \cdots & (1-a)(1-b) & (1-a)b &  a(1-b) & ab & \cdots & 0 \\
  &        &  & \cdots & &  &        &
\end{bmatrix}
}_{\K_{i,i-1}} 
\underbrace{
\begin{bmatrix}
\vdots \\
\x_{i-1}(s_1) \\
\x_{i-1}(s_2) \\
\x_{i-1}(s_3) \\
\x_{i-1}(s_4) \\
\vdots
\end{bmatrix}
}_{\x_{i-1}}
\end{tiny}
\end{equation}
which represents the bilinear interpolation of pixel $\x_i(s)$ from horizontally and vertically closest pixels to $\x_{i-1}(s+\v_i(s))$ in the previous frame ($\x_{i-1}(s_1)$, $\x_{i-1}(s_2)$, $\x_{i-1}(s_3)$, $\x_{i-1}(s_4)$) in a 2D image sequence as depicted in Figure~\ref{fig:K_example}.

The inaccuracy in (\ref{eqn:OFC_discrete}) due to the interpolation error, can be corrected with a residual term such as
\begin{align}
& \x_i = \K_{i,i-1}\x_{i-1} + \xr_i \\
\label{eqn:basicmotion2}
\Longrightarrow & \xr_i = \x_i - \K_{i,i-1}\x_{i-1}
\end{align}
where $\K_{i,i-1}=\K(\v_i)$ as stated earlier, $\xr_i$ is the residual signal after motion compensated estimation for $i$th frame. Reorganizing (\ref{eqn:basicmotion2}) we have
\begin{align}
\label{eqn:GenTV}
\begin{bmatrix}
\xr_2 \\
\vdots\\
\xr_{n_t}
\end{bmatrix} =& 
\begin{bmatrix} 
-\K_{2,1} & \I & \cdots & 0 \\
\vdots & & \ddots & \vdots \\
0 & \cdots &-\K_{n_t,n_t-1} & \I
\end{bmatrix} 
\begin{bmatrix}
\x_1 \\
\vdots\\
\x_{n_t}
\end{bmatrix} \\
\xr =& \Kt \x
\end{align}

Assuming $\xr$ in (\ref{eqn:GenTV}) is sparse, the signal can be reconstructed by the constrained minimization
\begin{equation}
\label{eqn:ObjFun_MTV}
\tilde{\x} = \argmin_{\x} \Vert\Kt\x\Vert_1 \text{   s.t.  } \Vert \y - \H\x \Vert^2_2 \leq  \epsilon
\end{equation}
It can be observed from  (\ref{eqn:GenTV}) and (\ref{eqn:ObjFun_MTV}) that the regularization term $\Vert\Kt\x\Vert_1$ is a more general form of temporal TV in (\ref{eqn:TV}) incorporating the motion information into the reconstruction, therefore we refer to it as "motion compensated total variation" or \textit{Motion-TV}. Note that $\xr_i$ in (\ref{eqn:basicmotion2}) will be significantly more sparse than the frame difference $\x_i - \x_{i-1}$ penalized by TV when dealing with large motion since it minimizes the pixel difference along the motion trajectory. The minimization in (\ref{eqn:ObjFun_MTV}) can be carried out with an Alternating Direction Method of Multipliers (ADMM) based approach with variable separation by solving the equivalent problem
\begin{align}
\nonumber \tilde{\x} = &\argmin_{\x} \Vert \p\Vert_1 \\
\label{eqn:ObjFun_MTV_ALAD}
&\text{  s.t.  } \p=\Kt\m,\m=\x,\Vert \y - \H\x \Vert^2_2 \leq  \epsilon
\end{align}
The steps of the ADMM algorithm at each iteration solving (\ref{eqn:ObjFun_MTV_ALAD}) can be summarized as
\begin{align}
\nonumber &\{\p, \m, \x\}  \gets \argmin\limits_{\p, \m,  \x} \Vert\p\Vert_1 + \mu_1\Vert\p - \Kt\m - \d_1\Vert_2^2\\
\nonumber  &\hspace{0.2\linewidth} + \mu_2\Vert \m - \x - \d_2 \Vert_2^2\\
                   &\hspace{0.2\linewidth} + \mu_3\Vert \y -\H\x - \s -\d_3 \Vert_2^2 \\
\label{eqn:upds}  &\s \gets \left\lbrace \begin{array}{ r } 
	0 \text{ if } \Vert\y-\H\x-\d_3 \Vert_2^2 \leq \epsilon\\
	\frac{\sqrt{\epsilon}(\y-\H\x-\d_3)}{\Vert\y-\H\x-\d_3\Vert_2}  \text{ if } \Vert\y-\H\x-\d_3 \Vert_2^2 > \epsilon
	\end{array}\right. \\
\label{eqn:updd1} &\d_1 \gets  \d_1 - (\p - \Kt\m) \\
\label{eqn:updd2} &\d_2 \gets  \d_2 - (\m - \x) \\
\label{eqn:updd3} &\d_3 \gets \d_3 - (\y -\H\x - \s)
\end{align}
in which $\mu_i$ are constants and $\d_i$ and $\s$ are intermediate variables. Details of the minimization algorithm are given in Algorithm~\ref{alg:admm} in which the function $\soft(.)$ is the element wise soft thresolding operation defined as
\begin{equation}
\label{eqn:soft_thr}
\soft(a, \tau) \triangleq \left\lbrace \begin{array}{ r }  (|a|-\tau)\dfrac{a}{|a|} \text{ if } |a| > \tau \\ 0 \text{ if } |a| \leq \tau \end{array}\right.
\end{equation}
A detailed derivation of the ADMM  approach to solve (\ref{eqn:ObjFun_MTV}) is out of the scope of this paper; readers who are interested are encouraged to read \cite{Goldstein2009}, \cite{Boyd2011}, \cite{Ramani2011} and the references therein.

\begin{algorithm} [!t]
\caption{ADMM minimization to solve (\ref{eqn:ObjFun_MTV})}
\label{alg:admm} 
\begin{algorithmic}[1] 
\Procedure{\textit{MotionTV\_ADMM}}{$\y,\H,\Kt,\epsilon$}
\Statex \textit{Minimizes $\Vert\p\Vert_1$ s.t $\p = \Kt\m$, $\m=\x$, $\Vert\y-\H\x\Vert_2^2\leq \epsilon$}
\State Set $\mu_1,\mu_2,\mu_3 > 0$, $\d_1\d_2,\d_3,\s = 0, \x=\H'\y, \m=\x$
\While{\textit{convergence criteria not met}}
	\State 
	$\p \gets  \argmin\limits_{\p}  \Vert\p\Vert_1 + \mu_1\Vert\p - \Kt\m - \d_1\Vert_2^2 $
	\Statex \hspace{0.2\linewidth} $ =	\soft( \Kt\m + \d_1, \frac{1}{2\mu_1})  $\label{alg:minp1}

	\State $\m \gets \argmin\limits_{\m} \begin{array}{l}
	\\
	\mu_1\Vert\p- \Kt\m - \d_1\Vert_2^2 \\
	 + \mu_2\Vert \m - \x - \d_2 \Vert_2^2
	 \end{array}$ \label{alg:minm1}
	\Statex \hspace{0.2\linewidth} $ = \left( \frac{\mu_2}{\mu_1} \I + \Kt'\Kt\right)^{-1} 
			\begin{array}{l} \\
			 \left[ \Kt'(\p-\d_1)  \right. \\
			  \left. + \frac{\mu_2}{\mu_1} (\x+\d_2) \right]
			  \end{array} $
	\Statex
	\State $\s \gets \left\lbrace \begin{array}{ r } 
	0 \text{ if } \Vert\y-\H\x-\d_3 \Vert_2^2 \leq \epsilon\\
	\frac{\sqrt{\epsilon}(\y-\H\x-\d_3)}{\Vert\y-\H\x-\d_3\Vert_2}  \text{ if } \Vert\y-\H\x-\d_3 \Vert_2^2 > \epsilon
	\end{array}\right.$
	\State $\x \gets  \argmin\limits_{\x} \begin{array}{l}
	\\
	\mu_2\Vert \m - \x - \d_2 \Vert_2^2 \\
	  + \mu_3\Vert \y -\H\x - \s -\d_3 \Vert_2^2
	 \end{array}$ \label{alg:minx1}
	 \Statex \hspace{0.2\linewidth} $ = \left( \frac{\mu_2}{\mu_3} \I + \H'\H\right)^{-1} 
	 									\begin{array}{l} \\ 
	 									\left[ \H'(\y-\s-\d_3)  \right. \\
	 									 \left. + \frac{\mu_2}{\mu_3} (\m-\d_2) \right]
	 									  \end{array} $
     \Statex
	 \State $\d_1 \gets \d_1 - (\p - \Kt\m)$
	 \State $\d_2 \gets \d_2 - (\m - \x)$
	 \State $\d_3 \gets \d_3 - (\y -\H\x - \s)$
\EndWhile
\State \textbf{return} $\x$
\EndProcedure 
\end{algorithmic} 
\end{algorithm}

The minimization in (\ref{eqn:ObjFun_MTV}) with \textit{Motion-TV} provides reconstruction without any use of reference frames and SNR penalty unlike (\ref{eqn:Residual}), however it still requires motion vectors. It is possible to estimate these motion vectors with a separate optimization from an initial low quality estimate of the signal or jointly with the signal itself.

\subsection{Separate Optimization of the Signal and the Motion}

In order to estimate the motion vectors, an initial reconstruction of the signal can be used which may suffer from artifacts such as temporal blurring but can still be sufficient for motion estimation with suitable estimation algorithms. Therefore the signal can be reconstructed with a 3 step algorithm as shown in Algorithm~\ref{alg:motiontv}. 

In the initial reconstruction phase the signal is reconstructed with regularization $\Vert\R_t\x\Vert_1$ which can be temporal TV, temporal wavelet or temporal Fourier transform (DFT), in case of a periodic signal. This initial reconstruction can then be used in the motion estimation step which can be performed by various registration or motion estimation algorithms almost all minimizing an objective function in the form as in line~\ref{algline:reg} of Algorithm~\ref{alg:motiontv}. The term $R(\v)$ regularizes the motion field if necessary while $\Vert \Kt \x^* \Vert^2_2$ enforces the optical flow constraint as stated earlier.  Note that, in order to enable accurate  motion estimation in step 2, the initial reconstruction should be sufficiently accurate so that the optical flow constraint is satisfied.

The algorithm for registration or motion estimation depends on the motion characteristics of the signal. In the simplest case, there can be global motion such as translational or rotational, which is seen in case of a moving camera in video applications or a moving object in medical imaging. The global motion parameters can be determined with iterative search algorithms to minimize $\Vert \Kt \x^* \Vert^2_2$ without the need for regularization ($R(\v)=0$) because it is an over-complete problem with few unknowns. 

A harder and more interesting problem arises when there is localized motion as well, such as in the case of multiple moving objects at different directions or speeds, or objects changing shape non-uniformly where the motion vector at each pixel may differ from its neighbor in general. Estimating the motion field where  all pixels are allowed to have different motion vectors is generally referred as the deformable registration problem. Unlike global motion, the unknowns in deformable registration is multiple times the number of pixels (an unknown motion vector with 2 or 3 dimensions for each pixel) and therefore it is an ill-posed problem which can only be solved with the help of regularization. To reduce the number of unknowns and yet allow local motion, one may assume each block of pixels have the same motion vector, as done in video compression. However, this block-wise motion representation is not very accurate for shape deformation often encountered in medical images, therefore every pixel can be represented with a different motion vector for local registration. 

Problems involving both global and local motion can be solved by first estimating the global motion parameters and then determining the local motion through deformable registration. Global motion parameters can be robustly estimated even if there are local deformations or noise due to the over-complete nature of the problem. After correcting for the global motion, the local motion can be estimated by iteratively solving
\begin{equation}
\label{eqn:regstep}
\v \gets \v + \argmin\limits_{\triangle\v} \lambda R(\v+\triangle\v) + \Vert \mathbf{K_t}(\v+\triangle\v) \x \Vert^2_2
\end{equation}
until convergence, in which $\triangle\v$ is the small update of the motion field with a known step size. Many popular deformable registration methods such as variants of Demon's Algorithm \cite{Thirion1996}, \cite{Pennec1999}, \cite{Vercauteren2009} or Finite Element Model (FEM) based methods \cite{Gee1994}, \cite{Samavati2009} solve (\ref{eqn:regstep}) efficiently yielding a small deformation of the images towards the reference images. Motion estimation methods used in video compression such as block matching can still be adjusted to estimate a motion vector for each pixel, however when the deformation is significant aforementioned deformable registration algorithms can be more accurate. The regularization term in these algorithms are usually selected to enforce the smoothness of the motion field gradient ($R(\v)=\sum\limits_i \Vert\nabla \v_i \Vert_2^2$ for Demon's Algorithm, combinations of first and second spatial gradient often used for FEM based methods). Diffeomorphism, i.e. the condition of the geometric transform or motion field being smooth and invertible, can also be enforced to avoid physically unlikely motion fields either through regularization or by replacing the addition ($\v+\triangle\v$) in (\ref{eqn:regstep}) with other operations as in  \cite{Vercauteren2009}. 

Both the global and the local motion estimation are robust against the artifacts presented by the initial reconstruction step either due to over-completeness or regularization of the motion field. After the registration step, the final reconstruction of the signal is carried out by enforcing the \textit{Motion-TV} prior, using the estimated motion.

\begin{algorithm} [!t]
\caption{\textit{Motion-TV} regularization with separate motion estimation}
\label{alg:motiontv} 
\begin{algorithmic}[1] 
\Procedure{\textit{MotionTV\_Separate}}{$\y,\H,\epsilon$}
\State $\x^* \gets \argmin\limits_{\x} \Vert\R_t\x\Vert_1 \text{   s.t.  } \Vert \y - \H\x \Vert^2_2 \leq  \epsilon$ 
\Statex \Comment{\textit{Initial Reconstruction }}
\State $\v^* \gets \argmin\limits_{\v} \lambda R(\v) + \Vert \Kt \x^* \Vert^2_2 $ \label{algline:reg}
\Statex \Comment{\textit{Motion Estimation / Registration }}
\State $\tilde{\x} \gets \argmin\limits_{\x} \Vert\K_{\mathbf{t}}(\v^*)\x\Vert_1 \text{   s.t.  } \Vert \y - \H\x \Vert^2_2 \leq  \epsilon$ 
\Statex \Comment{\textit{Final Reconstruction }}
\State \textbf{return} $\tilde{\x}$
\EndProcedure 
\end{algorithmic} 
\end{algorithm}

\subsection{Joint Optimization of the Signal and the Motion}
\label{sec:joint_estimation}
Instead of separately solving for $\x$ and $\v$, each involving a number of iterations, another approach can be estimating them jointly by alternating their iterations such as in
\begin{align}
\label{eqn:admmstep}
\nonumber &\{\p, \m, \x\} \gets  \argmin\limits_{\p, \m,  \x} \Vert\p\Vert_1 + \mu_1\Vert\p - \Kt\m - \d_1\Vert_2^2\\
\nonumber & \hspace{0.2\linewidth}+ \mu_2\Vert \m - \x - \d_2 \Vert_2^2 \\
 & \hspace{0.2\linewidth}+ \mu_3\Vert \y -\H\x - \s -\d_3 \Vert_2^2
 \end{align}
 \begin{align}
 \label{eqn:regstep2}
& \v \gets  \v + \argmin\limits_{\triangle\v} \lambda R(\v+\triangle\v) + \Vert \mathbf{K_t}(\v+\triangle\v) \x \Vert^2_2\\
&\upd  \s, \d_1, \d_2, \d_3 
\end{align}
where $\s, \d_1, \d_2, \d_3$ are updated the same as in (\ref{eqn:upds}), (\ref{eqn:updd1}), (\ref{eqn:updd2}) and (\ref{eqn:updd3}).
This attempt however fails miserably due to noisy artifacts in $\x$ during the iterations introduced by the term $\H'\y$ such as in  line~\ref{alg:minx1} of Algorithm~\ref{alg:admm}. $\H$ is an ill conditioned matrix as defined by the compressed sensing problem and the term $\H'\y=\H'(\H\x+\n)$ (or $\left(\frac{\mu_2}{\mu_3} \I + \H'\H\right)^{-1}\H'(\H\x+\n)$ as in ADMM based algorithms such as Algorithm~\ref{alg:admm}) results in noisy artifacts (possibly along both spatial and temporal dimensions) due to non-zero entries off the main diagonal of $\H'\H$ resulting from incompleteness. This term is unavoidable for any algorithm that solves (\ref{eqn:L1}), (\ref{eqn:L1unconst}) or (\ref{eqn:L1an_pri}) since it results from the derivative of the data consistency term ($\Vert\y-\H\x\Vert_2^2$). In compressed sensing, this noise is iteratively filtered in the sparse transform domain with non-linear methods (such as soft-thresholding in line~\ref{alg:minp} of Algorithm~\ref{alg:admm}) and often diminish exponentially, although not completely removed until convergence. Therefore estimating $\v$ from a noisy $\x$ by solving (\ref{eqn:regstep2}) before convergence leads to an incorrect motion field and a non-sparse $\Kt\x$ which in turn inhibits the convergence of both $\v$ and $\x$ or leads to incorrect reconstruction.

Based on this observation, we propose to filter $\x$ resulting from solving (\ref{eqn:admmstep}), and  solve $\v$ by replacing $\x$ with the filtered version $\x^*$ in (\ref{eqn:regstep2}). To filter $\x$, we again make use of the compressed sensing theory, which states that the noisy artifacts in the signal can be removed by non-linear filtering in an incoherent transform domain.  The noise in $\x$ can be removed by solving
\begin{equation}
\label{eqn:filterx}
\x^* = \argmin_{\hat{\x}} \beta\Vert \mathbf{\Phi_t}\hat{\x} \Vert_1 +  \Vert \x - \hat{\x} \Vert_2^2
\end{equation}
where $\beta$ can be adjusted at each iteration according to noise power in the sparse temporal transform domain defined by $\mathbf{\Phi_t}$. Provided that $\mathbf{\Phi_t}\mathbf{\Phi_t}'=\mathbf{\Phi_t}'\mathbf{\Phi_t}=\I$, (\ref{eqn:filterx}) can be minimized with a single step
\begin{equation}
\x^* = \mathbf{\Phi_t}'	\soft( \mathbf{\Phi_t}\x, \frac{\beta}{2})
\end{equation}
where $\soft(.)$ is an element-wise soft thresholding operator as defined earlier. $\mathbf{\Phi_t}$ must be selected as a sufficiently sparse orthonormal transform, a temporal transform such as temporal wavelets or DFT if applicable, in order to remove the noise at the expense of removing a part of the signal as well. The main advantage of using temporal wavelets or DFT as $\mathbf{\Phi_t}$ is that they separate the signal in the (temporal) frequency domain which usually leads to sparse high pass components for natural signals. As a result, the thresholding operation in this transform domain leads to a temporally blurred signal in the image domain, with blurring decreasing at every iteration as noise power and $\beta$ gets smaller. This however is not a significant problem for global or local registration steps due to the following reasons:
\begin{itemize}
\item Global registration parameters can still be estimated from temporally blurry image robustly even in the initial iterations due to being a largely over-complete problem. In fact global registration can be performed only in the first iteration and not necessarily repeated again. Small inaccuracies in the estimation of global parameters can be compensated with local motion.
\item The local registration step solving (\ref{eqn:regstep}) advances the vector field a step towards the motion direction at every iteration. Therefore the motion field is updated only considering large motion in the initial steps disregarding any small motion that is blurred out by the temporal blurring. As the iterations progress and the blurring is reduced, smaller vectors in the motion field can also be estimated. This is very similar to spatially scaling the images to provide multi-resolution registration and provides robustness towards correct estimation of the motion field.
\end{itemize}
Consequently, assuming that the initial motion field is already corrected for global motion, the signal and the motion can be estimated jointly by iteratively applying
\begin{align}
\nonumber &\{\p, \m, \x\} \gets  \argmin\limits_{\p, \m,  \x} \Vert\p\Vert_1 + \mu_1\Vert\p - \Kt\m - \d_1\Vert_2^2\\
\nonumber &\hspace{0.2\linewidth}+ \mu_2\Vert \m - \x - \d_2 \Vert_2^2\\
 &\hspace{0.2\linewidth}+ \mu_3\Vert \y -\H\x - \s -\d_3 \Vert_2^2 \\
 &\x^* \gets  \argmin_{\hat{\x}} \beta\Vert \mathbf{\Phi_t}\hat{\x} \Vert_1 +  \Vert \x - \hat{\x} \Vert_2^2\\
 &\v \gets  \v + \argmin\limits_{\triangle\v} \lambda R(\v+\triangle\v) + \Vert \mathbf{K_t}(\v+\triangle\v) \x^* \Vert^2_2\\
&\upd  \s, \d_1, \d_2, \d_3 
\end{align}
with $\beta \longrightarrow 0$ exponentially as iterations progress. Note that the above iterations essentially solve the following problem:
\begin{align}
\nonumber \tilde{\x},\tilde{\v} =& \argmin_{\x,\v} \Vert\K_{\mathbf{t}}(\v^*)\x\Vert_1+\lambda R(\v)+\Vert \Kt \x^* \Vert^2_2 \\
\label{eqn:ObjFun_MTV_joint} & \text{   s.t.  } \Vert \y - \H\x \Vert^2_2 \leq  \epsilon, \v^*=\v, \x^*=\x
\end{align}
The details of the algorithm to minimize (\ref{eqn:ObjFun_MTV_joint}) is given in Algorithm~\ref{alg:admm_joint}. Notice that with $\v$ and therefore $\Kt$ being a variable in the optimization, the constrained and unconstrained formulations (such as in (\ref{eqn:L1}) and (\ref{eqn:L1unconst}) respectively) are no longer equivalent, hence the constrained formulation in (\ref{eqn:ObjFun_MTV_joint}) is a necessity.

\begin{algorithm} [!t]
\caption{ADMM minimization to solve (\ref{eqn:ObjFun_MTV_joint})}
\label{alg:admm_joint} 
\begin{algorithmic}[1] 
\Procedure{\textit{Joint\_MotionTV\_ADMM}}{$\y,\H,\epsilon$}
\Statex \textit{Minimizes $\Vert\p\Vert_1+\lambda R(\v)+\Vert \Kt \x^* \Vert^2_2$}
\Statex \textit{ s.t $\p = \Kts\m$, $\m=\x$, $\x^*=\x$, $\v^*=\v$, $\Vert\y-\H\x\Vert_2^2\leq \epsilon$}
\State Set $\mu_1,\mu_2,\mu_3 > 0$, $\d_1\d_2,\d_3,\s = 0,  \alpha > 1$
\State Set $\x=\H'\y, \m=\x, \v=\v^*=0, \beta > 0$
\While{\textit{convergence criteria not met}}
	\State $\p \gets  \argmin\limits_{\p}  \Vert\p\Vert_1 + \mu_1\Vert\p - \Kts\m - \d_1\Vert_2^2 $
	\Statex \hspace{0.2\linewidth} $ =	\soft( \Kts\m + \d_1, \frac{1}{2\mu_1}) \label{alg:minp} $
	
	\State $\m \gets \argmin\limits_{\m} \begin{array}{l}
	\\
	\mu_1\Vert\p- \Kts\m - \d_1\Vert_2^2 \\
	 + \mu_2\Vert \m - \x - \d_2 \Vert_2^2
	 \end{array}$ \label{alg:minm2}
	\Statex \hspace{0.2\linewidth} $ = \left( \frac{\mu_2}{\mu_1} \I + \Kts'\Kts\right)^{-1} 
			\begin{array}{l} \\
			 \left[ \Kts'(\p-\d_1)  \right. \\
			  \left. + \frac{\mu_2}{\mu_1} (\x+\d_2) \right]
			  \end{array} $
	\Statex
	\State $\s \gets \left\lbrace \begin{array}{ r } 
	0 \text{ if } \Vert\y-\H\x-\d_3 \Vert_2^2 \leq \epsilon\\
	\frac{\sqrt{\epsilon}(\y-\H\x-\d_3)}{\Vert\y-\H\x-\d_3\Vert_2}  \text{ if } \Vert\y-\H\x-\d_3 \Vert_2^2 > \epsilon
	\end{array}\right.$
	\State $\x \gets  \argmin\limits_{\x} \begin{array}{l}
	\\
	\mu_2\Vert \m - \x - \d_2 \Vert_2^2 \\
	  + \mu_3\Vert \y -\H\x - \s -\d_3 \Vert_2^2
	 \end{array}$ \label{alg:minx2}
	 \Statex \hspace{0.2\linewidth}  $ = \left( \frac{\mu_2}{\mu_3} \I + \H'\H\right)^{-1} 
	 									\begin{array}{l} \\ 
	 									\left[ \H'(\y-\s-\d_3)  \right. \\
	 									 \left. + \frac{\mu_2}{\mu_3} (\m-\d_2) \right]
	 									  \end{array} $

	 \Statex
	 \State $\x^* \gets \argmin\limits_{\hat{\x}} \beta\Vert \mathbf{\Phi_t}\hat{\x} \Vert_1 +  \Vert \x - \hat{\x} \Vert_2^2$
	 \Statex \hspace{0.2\linewidth} $ = \mathbf{\Phi_t}'	\soft( \mathbf{\Phi_t}\x, \frac{\beta}{2})  $
	 \Statex
	 \State $\beta \gets \beta / \alpha$
	 \Statex
	 \State $\v \gets \v + \argmin\limits_{\triangle\v} \lambda R(\v+\triangle\v) + \Vert \mathbf{K_t}(\v+\triangle\v) \x^* \Vert^2_2$ \label{alg:minv}
	 \Statex
	 \State $\v^* \gets \v$
	 \Statex
	 \State $\d_1 \gets \d_1 - (\p - \Kts\m)$
	 \State $\d_2 \gets \d_2 - (\m - \x)$
	 \State $\d_3 \gets \d_3 - (\y -\H\x - \s)$
\EndWhile
\State \textbf{return} $\x$, $\v$
\EndProcedure 
\end{algorithmic} 
\end{algorithm}

\section{Experimental Results}
\label{sec:Results}

\begin{figure*}
\centering
\subfloat[][\label{fig:FullSampled1}Frames 1, 4, 7, 10, 12 of Dataset 1. The reconstruction results of the pixel locations indicated in frame 1 are shown in Figure~\ref{fig:Pixels1}]{
\includegraphics[scale=1]{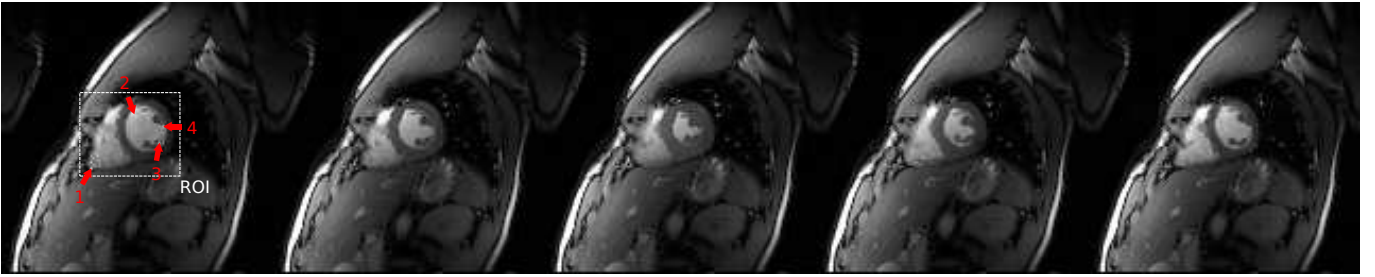}
}\\%
\subfloat[][\label{fig:FullSampled2}Frames 1, 4, 9, 11, 14 of Dataset 2. The reconstruction results of the pixel locations indicated in frame 1 are shown in Figure~\ref{fig:Pixels2}]{
\includegraphics[scale=.5]{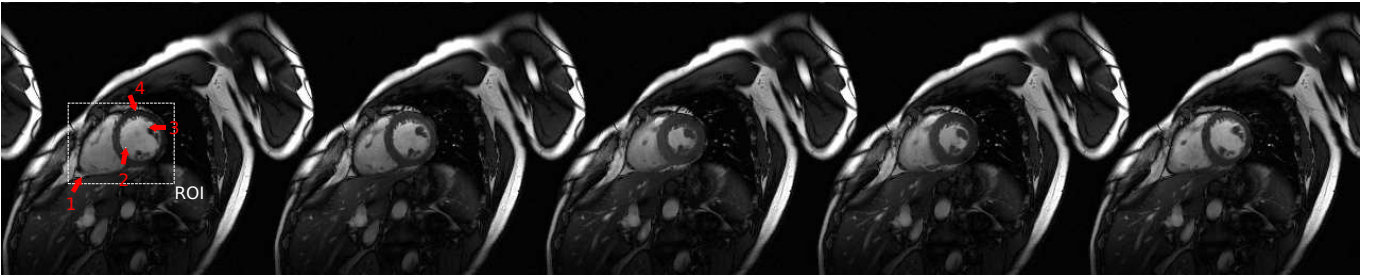}
}
\caption{Frames from fully sampled reconstructions of each dataset}%
\label{fig:FullSampled}%
\end{figure*}

The presented algorithms are tested with cardiac MRI datasets acquired with multiple coils to improve reconstruction quality. The measurements in parallel MRI can be formulated as
\begin{equation}
\y = \F \C \x + \n = \F \left[\C_{1}^T  \cdots \C_{N_c}^T\right]^T \x + \n
\end{equation} 
where $\F$ is the spatial Fourier transform, $\C_{i}, i=1,\ldots,N_c$ are the diagonal coil sensitivity matrices, $\y$ is the multi-coil measurement (k-space) data in time, $\n$ is the measurement noise and $\x$ is the image signal in time as defined earlier. In compressed sensing, the measurement is undersampled in a random fashion which can be represented as
\begin{equation}
\label{func:mri}
\y = \H\x + \n = \F_{u} \C \x + \n = \M \F \C \x + \n
\end{equation} 
where $\M$ is a subset of the rows of identity matrix and $\F_{u}=\M \F$ is the undersampled Fourier transform matrix. In cartesian MRI, undersampling is along the vertical and temporal dimensions only, fully sampling the horizontal dimension. 

Experiments are performed on 2 cardiac MRI cine datasets, acquired in steady state, i.e. measured signal intensity does not change in time. Each dataset is a 2D slice of a 3D volume and acquired over the duration of a single heart beat of a patient in time. Details on the datasets can be summarized as follows:
\begin{itemize}
\item \textit{Dataset 1:} 2D cardiac MRI dataset acquired on a 3T Siemens Trio scanner using a 32-coil matrix body array. Fully sampled data were acquired using a $128\times128$ matrix (FOV = $320\times320$ mm) and 22 temporal frames.
\item \textit{Dataset 2:} 2D cardiac MRI dataset acquired on a 3T Siemens Trio scanner using a 9-coil matrix body array. Fully sampled data were acquired using a $256\times256$ matrix (FOV = $320\times320$ mm) and 24 temporal frames.
\end{itemize}
For both of the datasets, the fully sampled data is retrospectively  by discarding a random subset of the samples along the vertical axis in each temporal frame, with varying subsampling patterns in different frames. The distribution of the selected samples is uniform along the temporal axis for any given vertical position and Gaussian along the vertical axis centered on zero frequency coefficients in spatial Fourier transform domain. Experiments are performed at subsampling rates R=8,10,12,and 14 (R=8 denotes $1/8$ of all the samples). The datasets are normalized so that the reconstruction of the fully sampled data leads to images with maximum intensity of 1. Sample frames from fully sampled reconstructions of the datasets can be seen in Figure~\ref{fig:FullSampled}.

The subsampled datasets are reconstructed with \textit{Motion-TV} both with separate and joint estimation of the motion the results of which are denoted by \textit{Motion-TV} and \textit{Joint Motion-TV} respectively in the figures. Note that at the experimented subsampling rates fully sampled reference frames to apply algorithms such as k-t FOCUSS are not applicable, therefore a comparison with these methods is not provided. Instead reconstruction results with temporal DFT and temporal TV regularization are presented for comparison. The difference of the first and last frame is included in both TV and \textit{Motion-TV} formulations due to the datasets being periodic. Optimization with temporal DFT and temporal TV are accomplished using Algorithm~\ref{alg:admm} replacing $\Kt$ with the proper operators. The steps involving matrix inversion in Algorithms \ref{alg:admm} and \ref{alg:admm_joint} are calculated with Conjugate Gradient algorithm as suggested in \cite{Ramani2011}. The parameters $\mu_i$ and $\epsilon$ in Algorithms~\ref{alg:admm} and \ref{alg:admm_joint} are determined empirically through simulations. It is possible to analytically determine $\beta$ and $\alpha$ through analysis on $\H$, however due to significant size of $\H$ in the example problems, these parameters are also decided empirically as $2$ and $1.09$ respectively. 

The datasets had no global motion but only local motion which was especially significant around the heart area, therefore the region around the heart shown in Figure~\ref{fig:FullSampled} is considered as the region of interest (ROI) when evaluating the reconstruction quality. Diffeo-symmetric Demon's registration algorithm with the recommended parameters in \cite{Vercauteren2009} is used as the deformable registration algorithm in order to estimate the motion in line~\ref{algline:reg} of Algorithm~\ref{alg:motiontv} as well as line~\ref{alg:minv} of Algorithm~\ref{alg:admm_joint}. The algorithm is applied without any multi-resolution scheme and is slightly modified to be applicable to complex valued datasets. The initial reconstruction step is accomplished with TV regularization for \textit{Motion-TV} while $\Phi_t$ is selected as temporal DFT for \textit{Joint Motion-TV}. Bilinear interpolation is used for motion compensation in formulating $\Kt$.

\begin{figure}
\centering
\subfloat[][\label{fig:RMSE1}Dataset 1]{
\includegraphics[scale=.5]{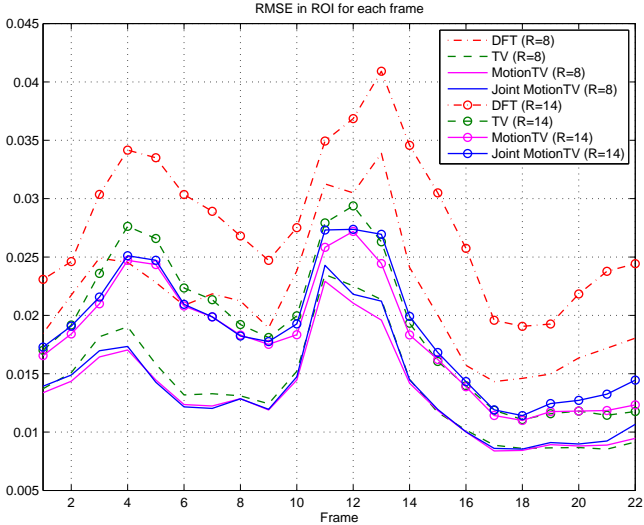}
}%
\qquad
\subfloat[][\label{fig:RMSE2}Dataset 2]{
\includegraphics[scale=.5]{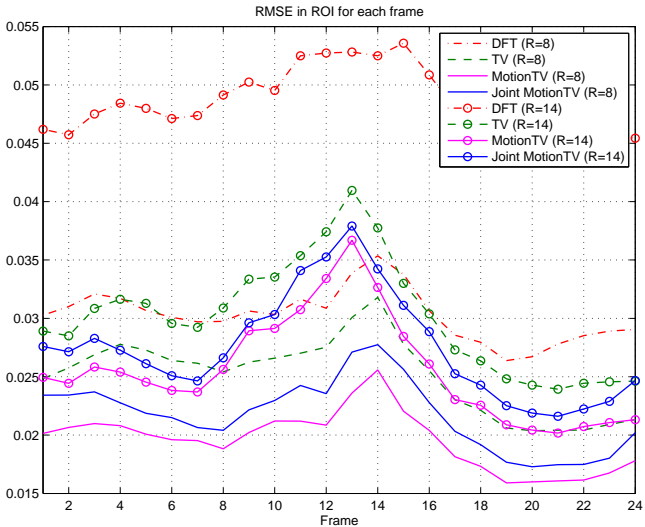}
}%
\caption{RMSE of the ROI in time for different datasets}%
\label{fig:RMSE}%
\end{figure}

\begin{figure*}
\centering
\subfloat[][\label{fig:RecSample1}Dataset 1]{
\includegraphics[scale=1]{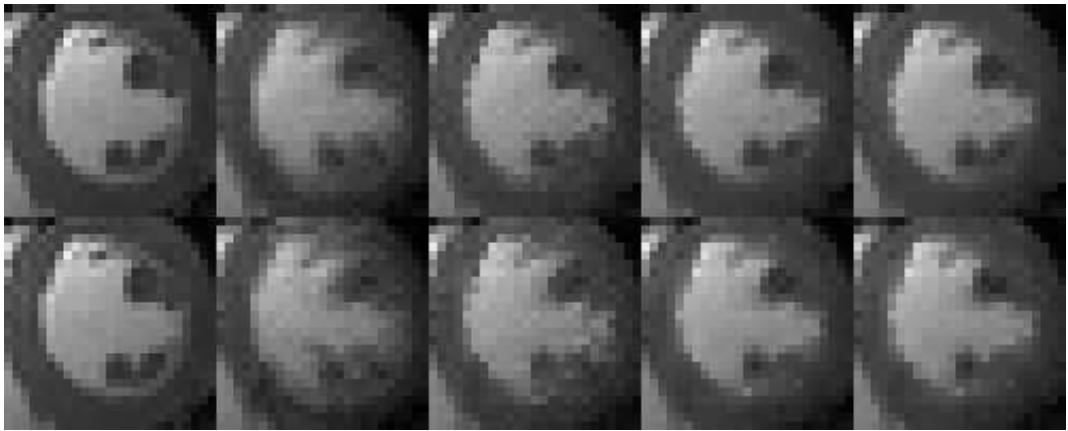}
}\\%
\subfloat[][\label{fig:RecSample2}Dataset 2]{
\includegraphics[scale=1]{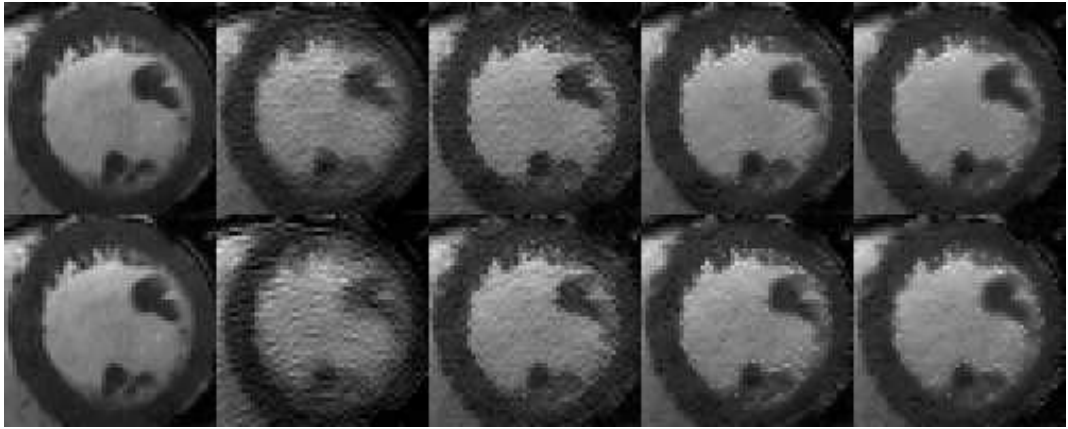}
}
\caption{Reconstructed samples from all datasets (frame 4). In all figures, from left to right, fully sampled, DFT, TV, \textit{Motion-TV} and \textit{Joint Motion-TV} reconstruction is shown. First row is the reconstruction at subsampling rate R=8 and the second row is at rate R=14.}%
\label{fig:RecSample}%
\end{figure*}

\begin{figure*}
\centering
\subfloat[][\label{fig:MVSample1_wtv8}Dataset 1, \textit{Motion-TV}, R=8]{
\includegraphics[scale=.25]{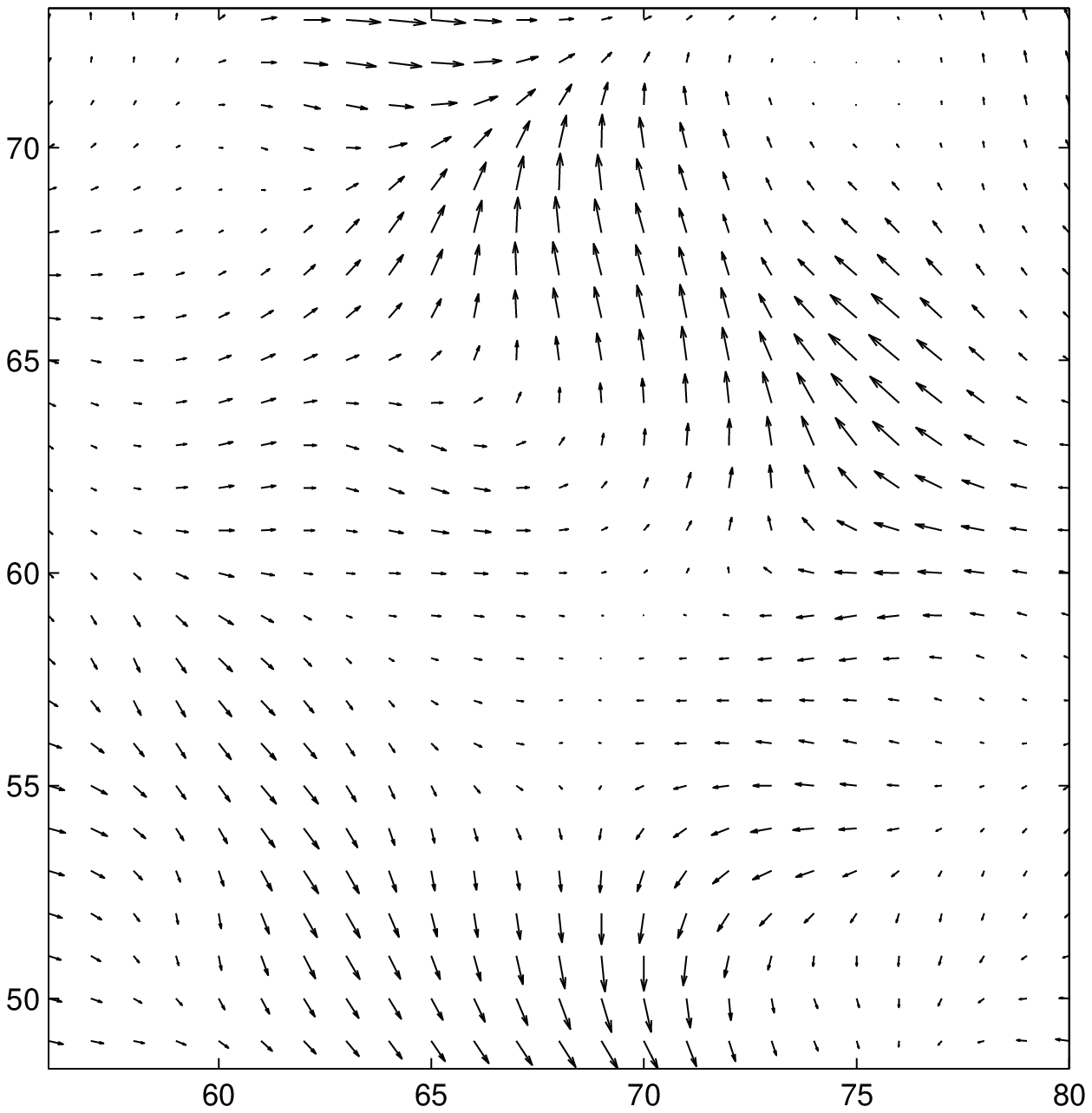}
}
\qquad
\subfloat[][\label{fig:MVSample1_awtv8}Dataset 1, \textit{Joint Motion-TV}, R=8]{
\includegraphics[scale=.25]{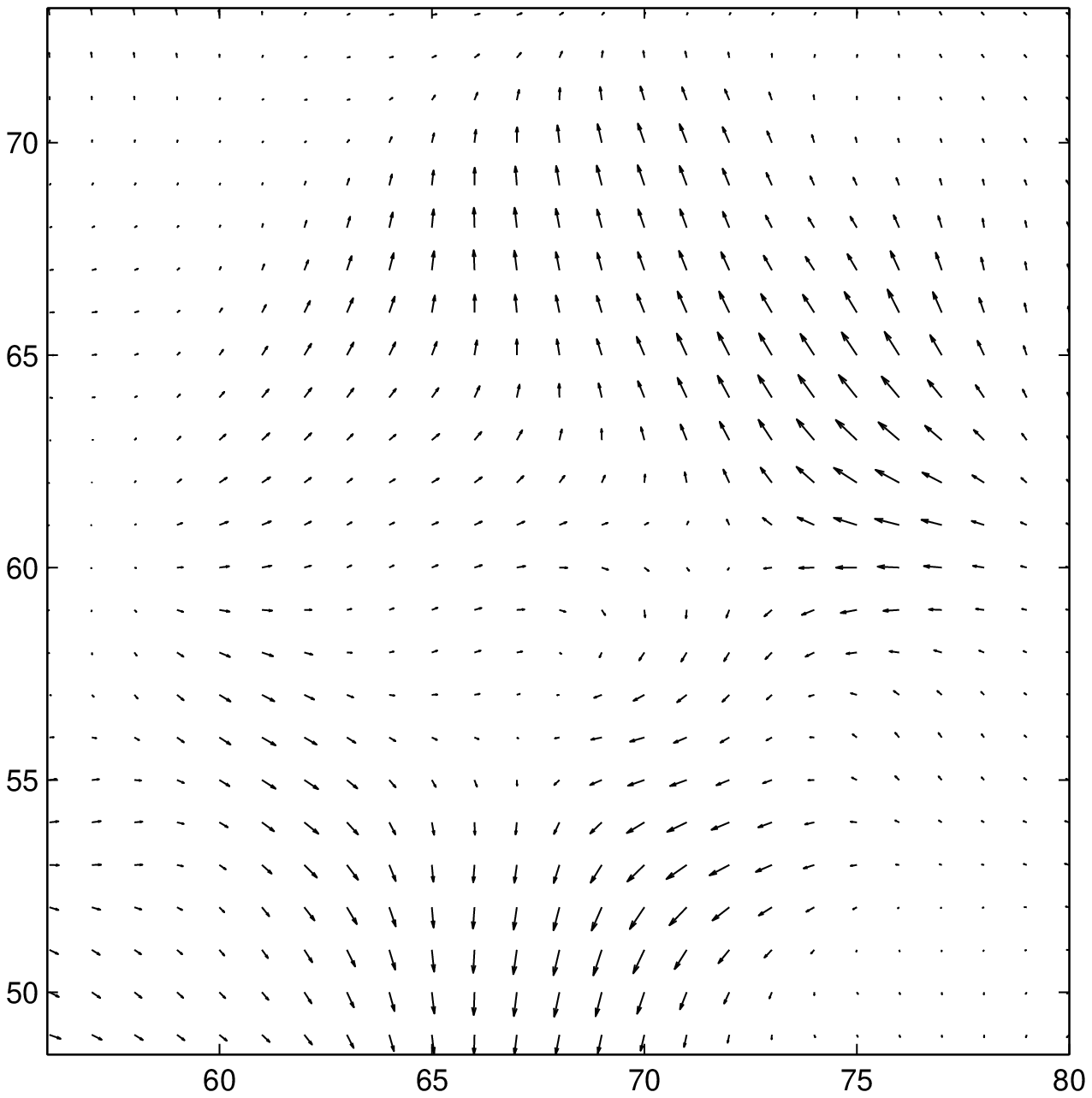}
}
\qquad
\subfloat[][\label{fig:MVSample1_wtv14}Dataset 1, \textit{Motion-TV}, R=14]{
\includegraphics[scale=.25]{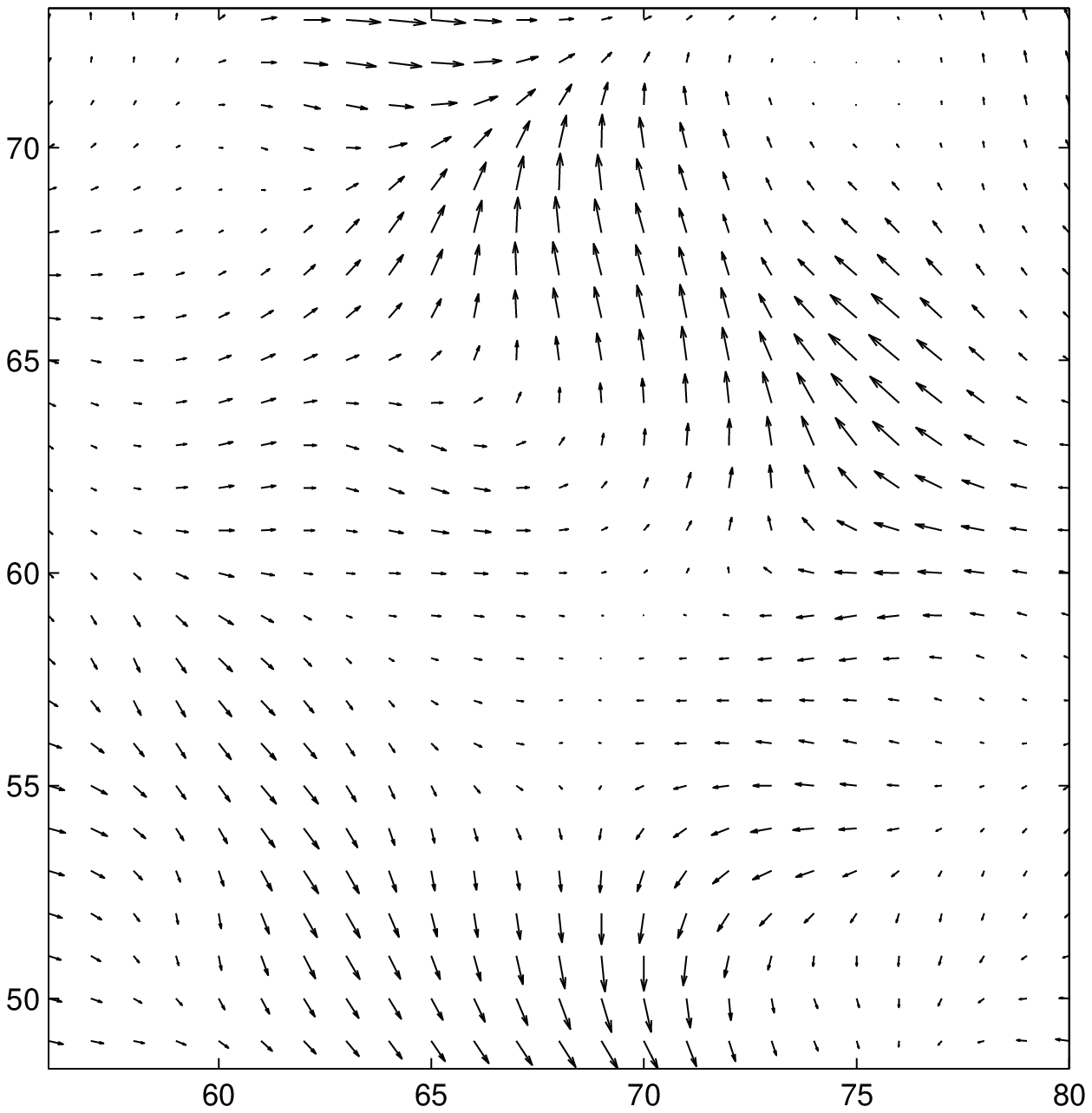}
}
\qquad
\subfloat[][\label{fig:MVSample1_awtv14}Dataset 1, \textit{Joint Motion-TV}, R=14]{
\includegraphics[scale=.25]{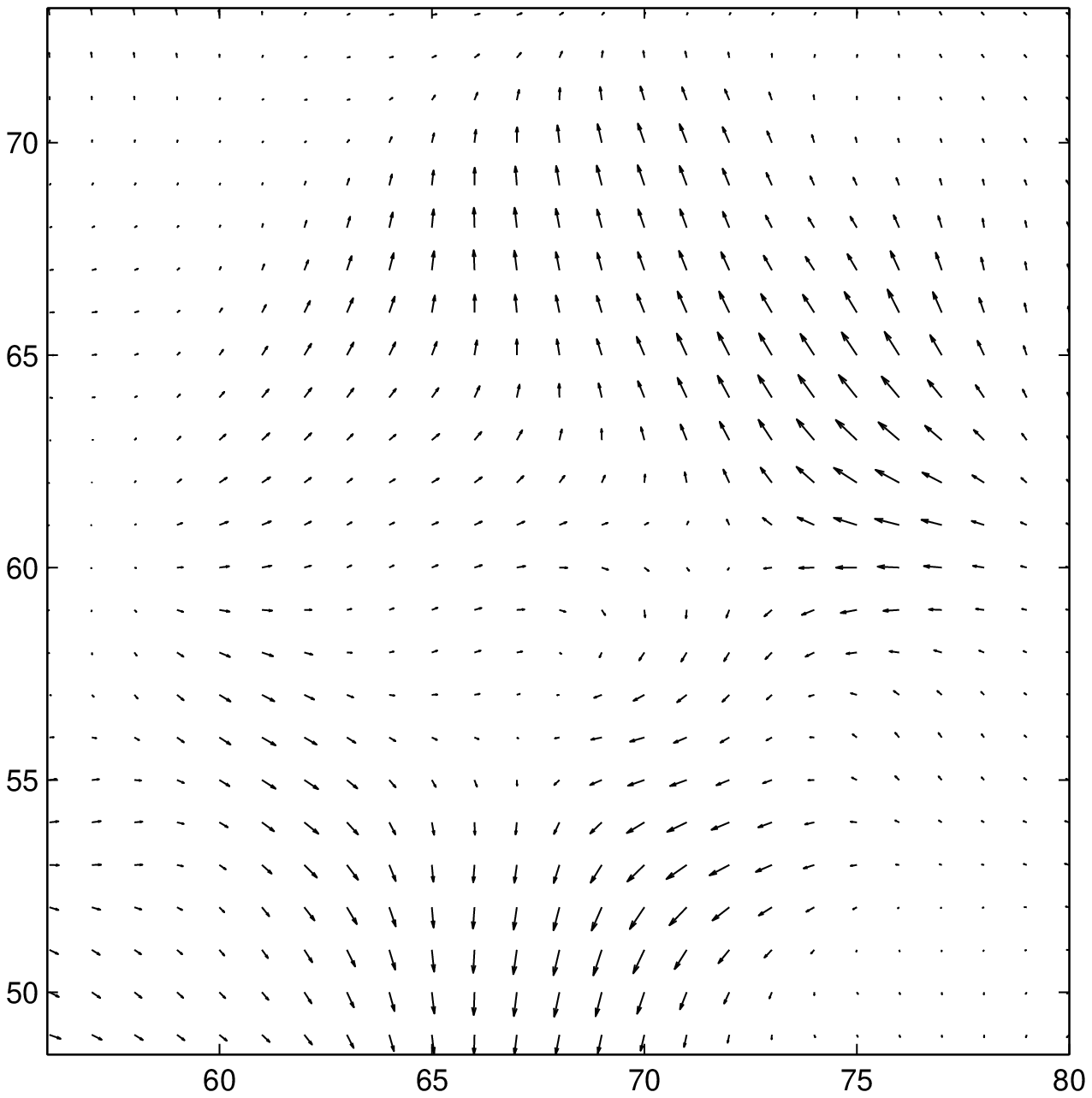}
}%
\\
\subfloat[][\label{fig:MVSample2_wtv8}Dataset 2, \textit{Motion-TV}, R=8]{
\includegraphics[scale=.25]{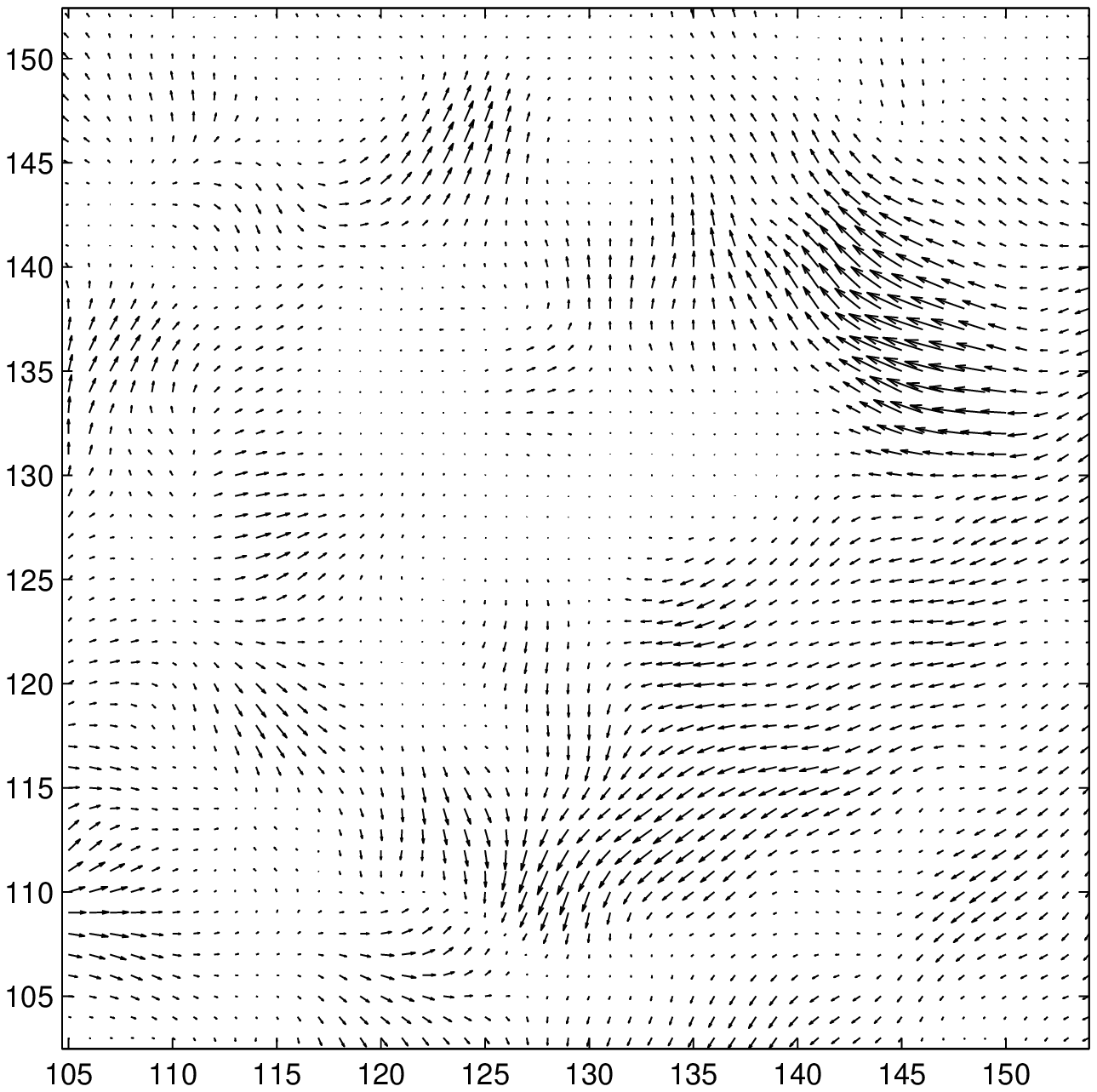}
}
\qquad
\subfloat[][\label{fig:MVSample2_awtv8}Dataset 2, \textit{Joint Motion-TV}, R=8]{
\includegraphics[scale=.25]{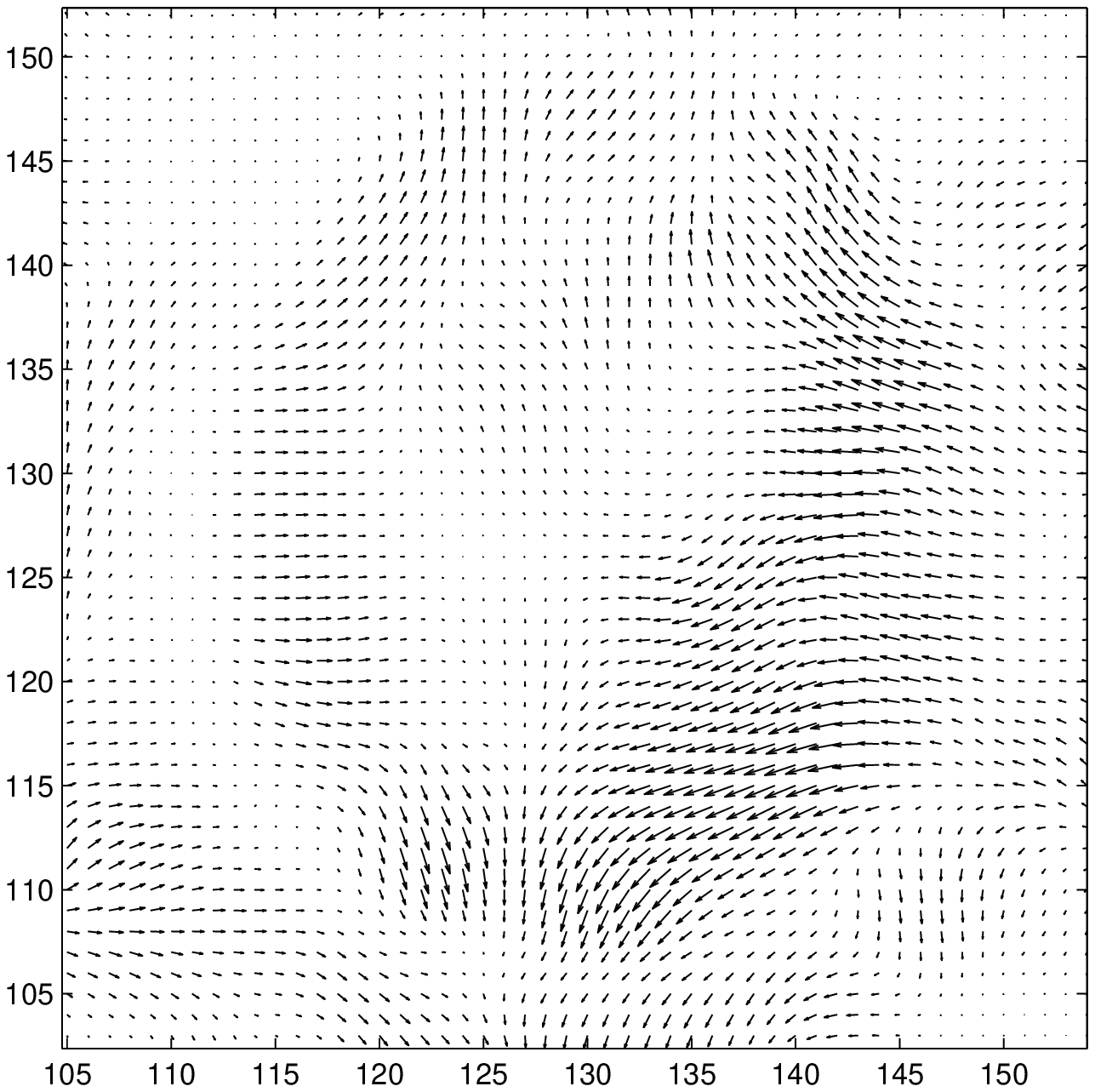}
}
\qquad
\subfloat[][\label{fig:MVSample2_wtv14}Dataset 2, \textit{Motion-TV}, R=14]{
\includegraphics[scale=.25]{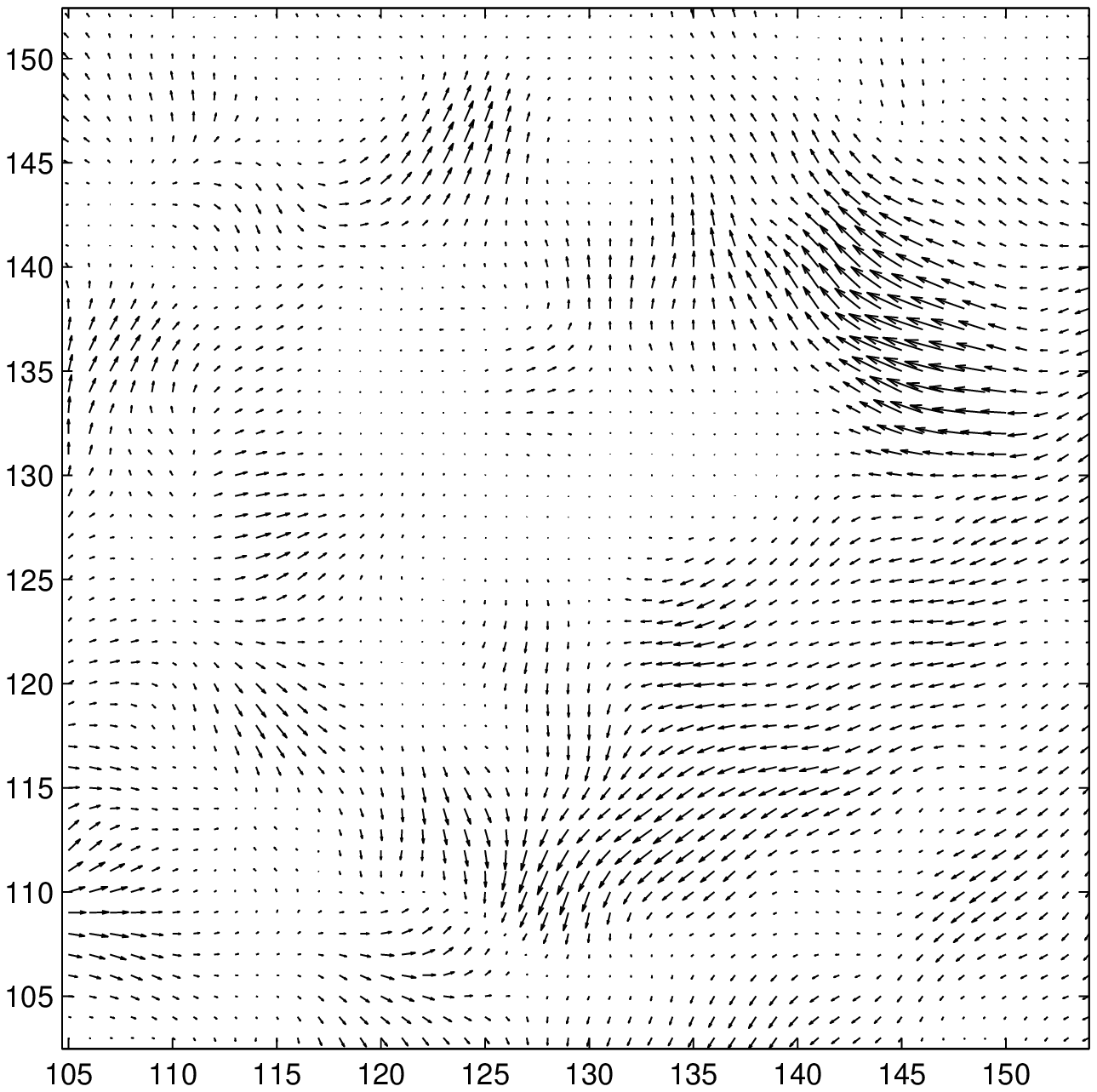}
}
\qquad
\subfloat[][\label{fig:MVSample2_awtv14}Dataset 2, \textit{Joint Motion-TV}, R=14]{
\includegraphics[scale=.25]{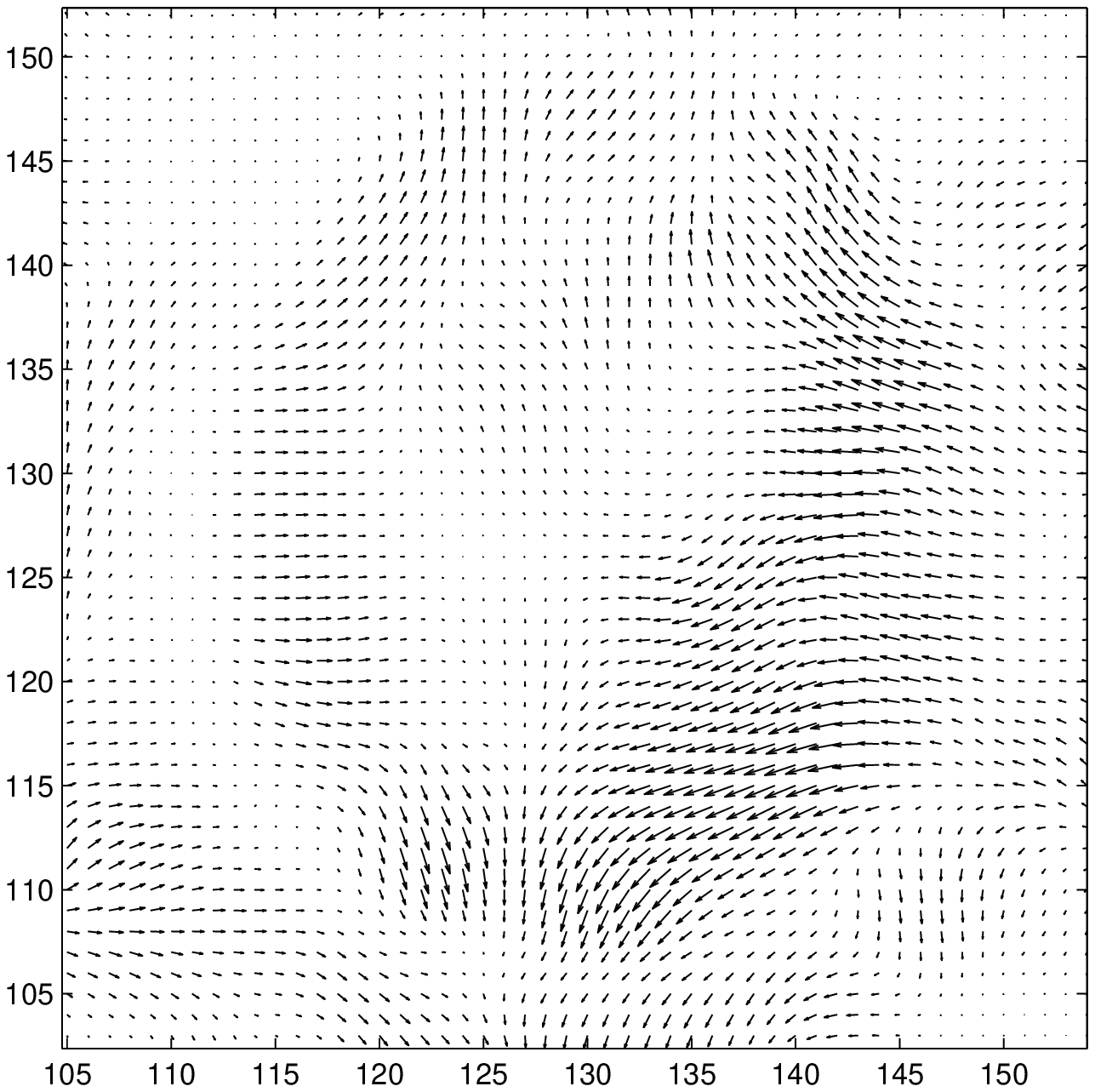}
}%
\caption{Estimated motion vectors by \textit{Motion-TV} and \textit{Joint Motion-TV} for all datasets (frame 4, regions shown in Figure~\ref{fig:RecSample}) for subsampling rates R=8 and R=14.}%
\label{fig:MVSample}%
\end{figure*}

The root mean squared error (RMSE) in the ROI of each frame with respect to fully sampled reconstructions are plotted in Figure~\ref{fig:RMSE} while the overall RMSE is listed in Tables~\ref{tab:RMSE1} and \ref{tab:RMSE2}. It can be seen that the RMSE improvement with respect to TV reconstruction is small in Dataset 1 but is quite significant in Dataset 2. This can be explained by the fact that the magnitude of motion vectors are much smaller in Dataset 1 and mostly not even greater than a single pixel (can be seen in Figure~\ref{fig:MVSample}). Therefore the benefit of \textit{Motion-TV} is limited. As the resolution and the scale of motion increase, the benefit of \textit{Motion-TV} and \textit{Joint Motion-TV} over TV and DFT can be seen more clearly in terms of RMSE in Figure~\ref{fig:RMSE2} and Table~\ref{tab:RMSE2}. The lower sparsity of temporal TV and DFT manifest as spatial noise or temporal blurring which can be observed in the samples shown in Figure~\ref{fig:RecSample}.

\begin{figure*}
\centering
\subfloat[][\label{fig:Pixels1}Dataset 1]{
\includegraphics[scale=.75]{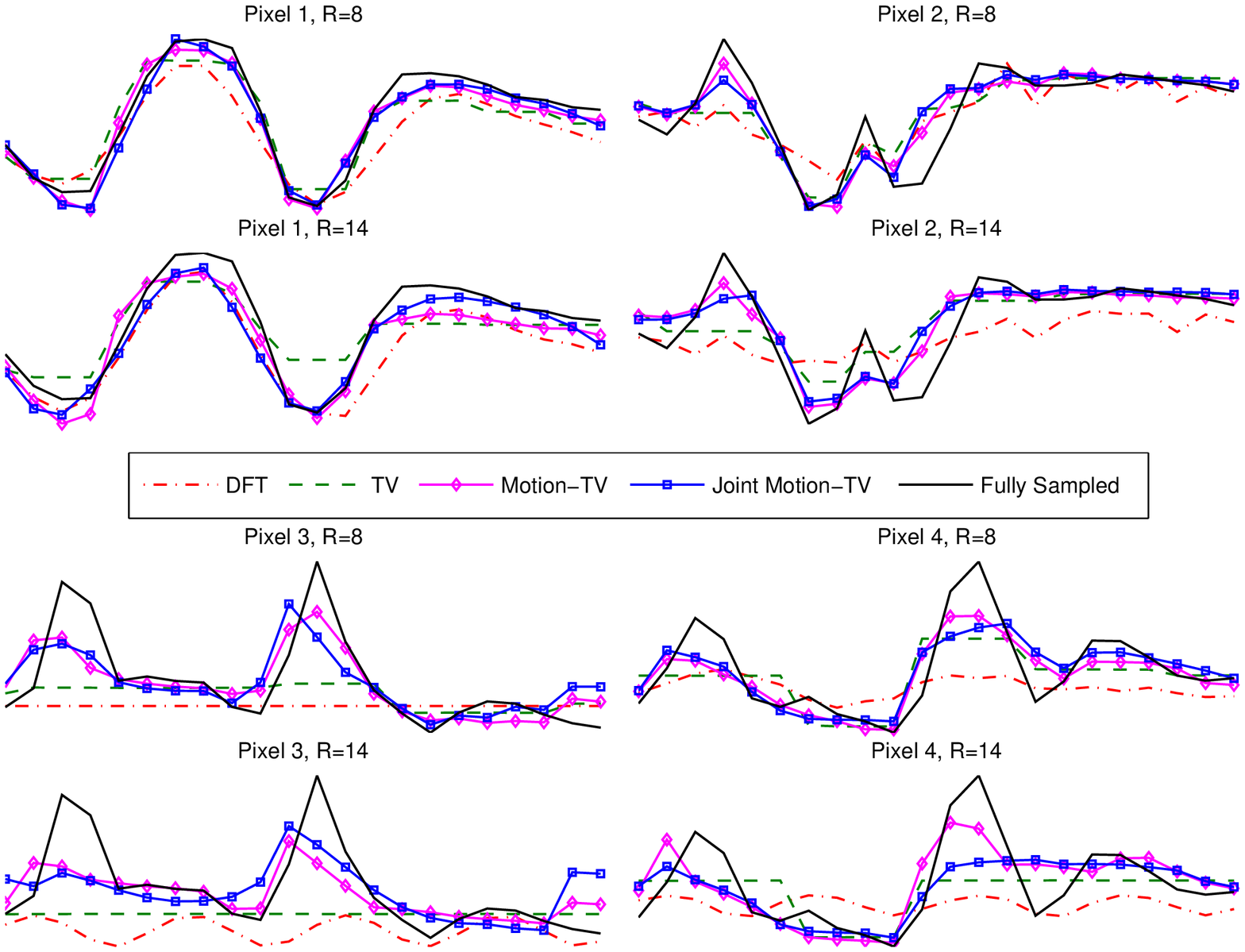}
}\\%
\subfloat[][\label{fig:Pixels2}Dataset 2]{
\includegraphics[scale=.75]{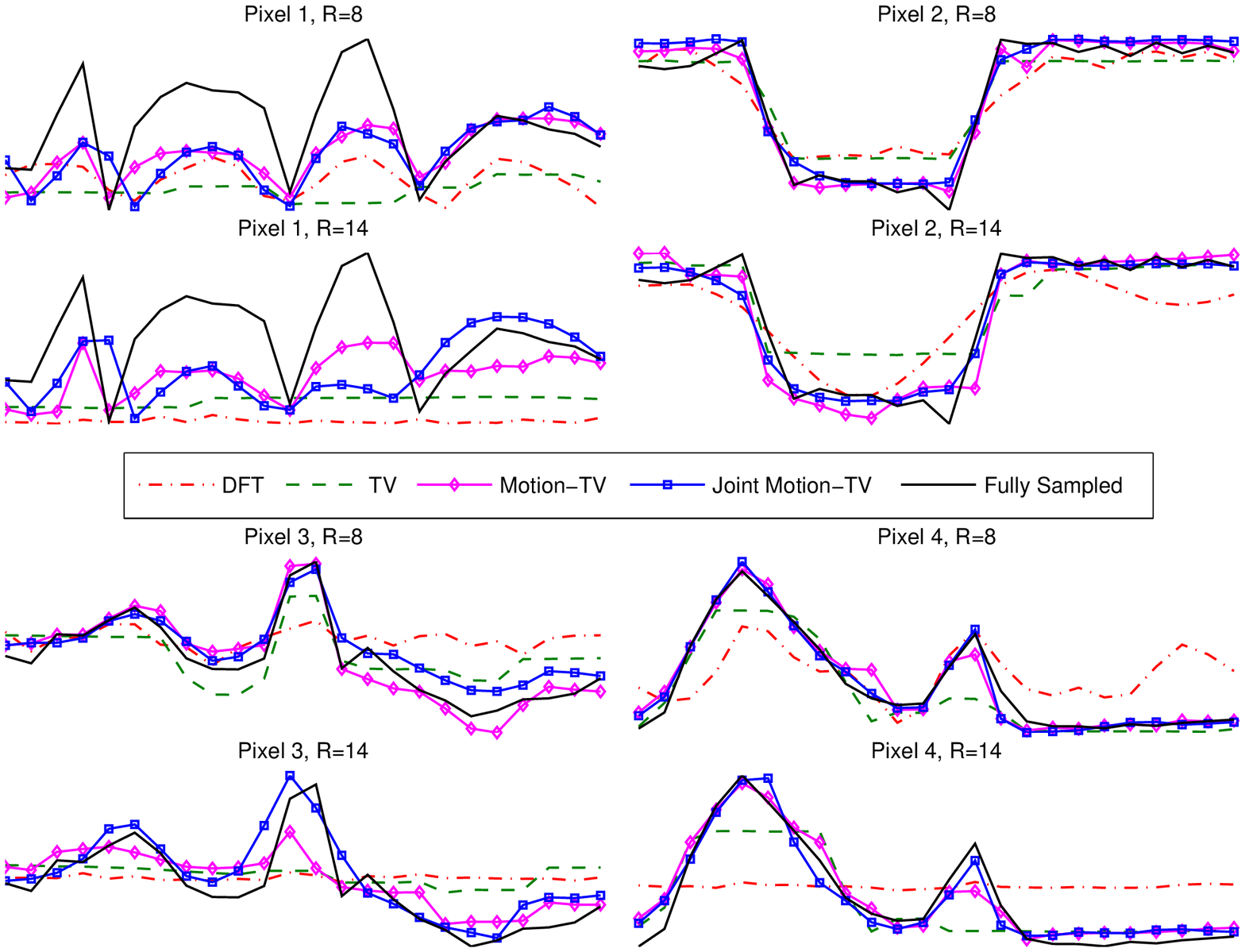}
}\\%
\caption{Change of intensity in time for the pixels shown in Figure~\ref{fig:FullSampled}.}%
\end{figure*}

The temporal change in the signal and its estimation accuracy is shown in Figures~\ref{fig:Pixels1} and \ref{fig:Pixels2} for the sample pixels numbered in Figure~\ref{fig:FullSampled} which are selected from most temporally varying areas in the Datasets. The temporal smoothing in DFT regularization can be very clearly seen in these pixel examples, although it is a commonly used sparsifying transform in compressed sensing dynamic MRI \cite{Otazo2010}. The TV regularization can perform better however still the staircase effect of TV is visible and the performance is mostly worse than \textit{Motion-TV} and \textit{Joint Motion-TV} which enforce sparsity in the temporal signal along the motion trajectory.

%


In the presented experiments, the performance of \textit{Motion-TV} was better than \textit{Joint Motion-TV} in general. We believe this is due to the larger number of parameters to be adjusted in the \textit{Joint Motion-TV} algorithm and a closer performance to \textit{Motion-TV} is possible with better optimization of parameters. This claim is also supported by the resulting motion fields shown in in Figure~\ref{fig:MVSample}, in which the field estimated by \textit{Joint Motion-TV} can be observed to be at least as consistent as the field estimated by \textit{Motion-TV}. In fact the motion field is more continuous along the tissue boundaries (edges) thanks to the multi-resolution effect during the estimation of motion field as discussed in Section~\ref{sec:joint_estimation}. Also note that when much larger motion is involved, the quality of the initial reconstruction in \textit{Motion-TV}  would suffer from severe artifacts or noise and the motion estimation step might therefore fail to estimate an accurate motion field. In these cases further iterations between motion estimation and signal reconstruction can be used which in turn leads to very long reconstruction time. The joint estimation however gradually estimates the motion field and is more robust and preferable in high motion cases for its efficient computation.


Each of the separate and joint \textit{Motion-TV} has benefits with regards to the other which can be summarized as follows:
\begin{itemize}
\item \textit{Separate Optimization: }Each of the optimization steps (initial estimation, motion estimation, final estimation) can be performed independently from each other with separate algorithms and/or implementations if necessary. This provides ease of implementation as well as robustness. It is also possible to correct the possible errors in the initial reconstruction or motion estimation steps before the final estimation (noise smoothed out or motion vectors manually corrected). The resulting reconstruction quality suffers if the initial reconstruction is not accurate enough for motion estimation. The overall reconstruction time is larger (close to twice with our implementation) than joint optimization case 
\item \textit{Joint Optimization: }Efficient in terms of execution time and motion estimation although some extra parameters need to be determined ($\beta$, $\alpha$) to ensure convergence and speed. The performance does not depend significantly on the quality of initial reconstructions since the signal and the motion are gradually estimated in joint iterations. Therefore joint reconstruction is preferable over separate reconstruction in case of signals with very large motion.
\end{itemize}

In terms of algorithm complexity, thanks to $\Kt$ being a very sparse matrix, line~\ref{alg:minm1} of Algorithm~\ref{alg:admm} can be efficiently calculated using an algorithm such as conjugate gradient method without a significant overhead (less than 10\%) on the total execution time of the algorithm when compared with simpler transforms such as temporal TV and temporal DFT. This is also as a result of having a very large $\H$, and the operations on $\H$ being the dominant factor in the speed of the minimization. The speed of convergence is also observed to be similar. For \textit{Joint Motion-TV}, the registration step (line~\ref{alg:minv} of Algorithm~\ref{alg:admm_joint}) adds around 20\% increase in the total execution time in our implementation, while the speed of convergence in terms of number of iterations is similar to the other algorithms. For all minimizations, 100 iterations are observed to be sufficient for convergence.

\begin{table}
\label{tab:RMSE}
\caption{RMSE for all frames of Dataset 1}
\centering
\begin{tabular}{l|cccc}
\label{tab:RMSE1}
& R=8 & R=10 & R=12 & R=14 \\
\hline
DFT & 0.0220 & 0.0239 & 0.0271 & 0.0286\\
TV   & 0.0146 & 0.0168 & 0.0187 & 0.0199\\
\textit{Motion-TV} & \textbf{0.0139} & \textbf{0.0158} & \textbf{0.0177} & \textbf{0.0187}\\
\textit{Joint Motion-TV} & 0.0143 & 0.0164 & 0.0183 & 0.0194
\end{tabular}
\caption{RMSE for all frames of Dataset 2}
\begin{tabular}{l|cccc}
\label{tab:RMSE2}
& R=8 & R=10 & R=12 & R=14\\
\hline
DFT & 0.0303 & 0.0379 & 0.0410 & 0.0485\\
TV   & 0.0253 & 0.0265 & 0.0288 & 0.0305\\
\textit{Motion-TV} & \textbf{0.0197} & \textbf{0.0214} & \textbf{0.0237} & \textbf{0.0260}\\
\textit{Joint Motion-TV} & 0.0219 & 0.0235 & 0.0256 & 0.0278
\end{tabular}
\end{table}

\section{Conclusion}
\label{sec:Conc}
A new regularization and reconstruction scheme is proposed for motion compensated compressed sensing to be used in reconstruction of images or volumes of moving objects in time. The proposed method can be used in scenarios that earlier motion compensation schemes are not applicable such as when no reference frames are present and outperforms known regularization methods without motion compensation when the motion is significant. In addition to improved regularization with a known motion field, it is also shown that, although not straightforward, it is possible to jointly estimate the motion and the signal efficiently by utilizing the principles of compressed sensing.

While the provided experiment results demonstrate the performance in cardiac MRI reconstruction, the proposed algorithms are by no means limited to this application and can be used in applications such as reconstruction of free breathing medical imaging data or correction of patient movement during compressed sensing reconstruction. The proposed algorithms can also be improved by making use of better interpolation methods or bi-directional motion estimation therefore reducing estimation error and increasing sparsity which will be investigated as a future work. 

\section{Acknowledgements}
The authors would like to thank Li Feng, Daniel Kim, Ricardo Otazo and Daniel K. Sodickson from NYU Medical Center for their help and support as well as for providing the cardiac MRI datasets.




%
%
%
\bibliographystyle{IEEEtran_nourl}
\bibliography{libupdate.bib}

\begin{thebibliography}{10}
\def\url#1{}
\csname url@samestyle\endcsname
\providecommand{\newblock}{\relax}
\providecommand{\bibinfo}[2]{#2}
\providecommand{\BIBentrySTDinterwordspacing}{\spaceskip=0pt\relax}
\providecommand{\BIBentryALTinterwordstretchfactor}{4}
\providecommand{\BIBentryALTinterwordspacing}{\spaceskip=\fontdimen2\font plus
\BIBentryALTinterwordstretchfactor\fontdimen3\font minus
  \fontdimen4\font\relax}
\providecommand{\BIBforeignlanguage}[2]{{%
\expandafter\ifx\csname l@#1\endcsname\relax
\typeout{** WARNING: IEEEtran.bst: No hyphenation pattern has been}%
\typeout{** loaded for the language `#1'. Using the pattern for}%
\typeout{** the default language instead.}%
\else
\language=\csname l@#1\endcsname
\fi
#2}}
\providecommand{\BIBdecl}{\relax}
\BIBdecl

\bibitem{Candes2008}
\BIBentryALTinterwordspacing
E.~Candes and M.~Wakin, ``{An Introduction To Compressive Sampling},''
  \emph{IEEE Signal Processing Magazine}, vol.~25, no.~2, pp. 21--30, Mar.
  2008.
  \url{http://ieeexplore.ieee.org/lpdocs/epic03/wrapper.htm?arnumber=4472240}
\BIBentrySTDinterwordspacing

\bibitem{Lustig2008}
\BIBentryALTinterwordspacing
M.~Lustig, D.~Donoho, J.~Santos, and J.~Pauly, ``{Compressed Sensing MRI},''
  \emph{IEEE Signal Processing Magazine}, vol.~25, no.~2, pp. 72--82, Mar.
  2008.
  \url{http://ieeexplore.ieee.org/lpdocs/epic03/wrapper.htm?arnumber=4472246}
\BIBentrySTDinterwordspacing

\bibitem{Liu2008}
\BIBentryALTinterwordspacing
B.~Liu, Y.~M.~Y. Zou, and L.~Ying, ``{SparseSENSE: application of compressed
  sensing in parallel MRI},'' in \emph{Information Technology and Applications
  in Biomedicine, 2008. ITAB 2008. International Conference on}, vol.~2,
  no.~3.\hskip 1em plus 0.5em minus 0.4em\relax Ieee, May 2008, pp. 127--130.
  \url{http://ieeexplore.ieee.org/lpdocs/epic03/wrapper.htm?arnumber=4570588
  http://ieeexplore.ieee.org/xpls/abs\_all.jsp?arnumber=4570588}
\BIBentrySTDinterwordspacing

\bibitem{Yu2009}
\BIBentryALTinterwordspacing
H.~Yu and G.~Wang, ``{Compressed sensing based interior tomography},''
  \emph{Physics in medicine and biology}, vol.~54, p. 2791, 2009.
  \url{http://iopscience.iop.org/0031-9155/54/9/014}
\BIBentrySTDinterwordspacing

\bibitem{Provost2009}
\BIBentryALTinterwordspacing
J.~Provost and F.~Lesage, ``{The application of compressed sensing for
  photo-acoustic tomography.}'' \emph{IEEE transactions on medical imaging},
  vol.~28, no.~4, pp. 585--94, Apr. 2009.
  \url{http://www.pubmedcentral.nih.gov/articlerender.fcgi?artid=3169509\&tool=pmcentrez\&rendertype=abstract}
\BIBentrySTDinterwordspacing

\bibitem{Chen2009}
\BIBentryALTinterwordspacing
G.-H. Chen, J.~Tang, and J.~Hsieh, ``{Temporal resolution improvement using
  PICCS in MDCT cardiac imaging},'' \emph{Medical Physics}, vol.~36, no.~6, p.
  2130, 2009.  \url{http://link.aip.org/link/MPHYA6/v36/i6/p2130/s1\&Agg=doi}
\BIBentrySTDinterwordspacing

\bibitem{Gamper2008}
\BIBentryALTinterwordspacing
U.~Gamper, P.~Boesiger, and S.~Kozerke, ``{Compressed sensing in dynamic
  MRI.}'' \emph{Magnetic resonance in medicine : official journal of the
  Society of Magnetic Resonance in Medicine / Society of Magnetic Resonance in
  Medicine}, vol.~59, no.~2, pp. 365--73, Feb. 2008.
  \url{http://www.ncbi.nlm.nih.gov/pubmed/18228595}
\BIBentrySTDinterwordspacing

\bibitem{Otazo2010}
\BIBentryALTinterwordspacing
R.~Otazo, D.~Kim, L.~Axel, and D.~K. Sodickson, ``{Combination of compressed
  sensing and parallel imaging for highly accelerated first-pass cardiac
  perfusion MRI.}'' \emph{Magnetic resonance in medicine : official journal of
  the Society of Magnetic Resonance in Medicine / Society of Magnetic Resonance
  in Medicine}, vol.~64, no.~3, pp. 767--76, Sep. 2010.
  \url{http://www.pubmedcentral.nih.gov/articlerender.fcgi?artid=2932824\&tool=pmcentrez\&rendertype=abstract}
\BIBentrySTDinterwordspacing

\bibitem{Usman2011}
\BIBentryALTinterwordspacing
M.~Usman, C.~Prieto, T.~Schaeffter, and P.~G. Batchelor, ``{k-t group sparse: A
  method for accelerating dynamic MRI.}'' \emph{Magnetic resonance in medicine
  : official journal of the Society of Magnetic Resonance in Medicine / Society
  of Magnetic Resonance in Medicine}, vol. 1176, pp. 1163--1176, Mar. 2011.
  \url{http://www.ncbi.nlm.nih.gov/pubmed/21394781}
\BIBentrySTDinterwordspacing

\bibitem{Vaswani2008}
\BIBentryALTinterwordspacing
N.~Vaswani, ``{Kalman filtered compressed sensing},'' in \emph{Image
  Processing, 2008. ICIP 2008. 15th IEEE International Conference on},
  no.~1.\hskip 1em plus 0.5em minus 0.4em\relax IEEE, 2008, pp. 893--896.
  \url{http://ieeexplore.ieee.org/xpls/abs\_all.jsp?arnumber=4711899}
\BIBentrySTDinterwordspacing

\bibitem{Lam2010}
F.~Lam, D.~Hernando, K.~F. King, and Z.-P. Liang, ``{Compressed Sensing
  Reconstruction in the Presence of a Reference Image},'' in \emph{Proceedings
  18th Scientific Meeting International Society for Magnetic Resonance in
  Medicine}, vol.~18, 2010, pp. 4861--4861.

\bibitem{Jung2009a}
\BIBentryALTinterwordspacing
H.~Jung, K.~Sung, K.~S. Nayak, E.~Y. Kim, and J.~C. Ye, ``{k-t FOCUSS: a
  general compressed sensing framework for high resolution dynamic MRI.}''
  \emph{Magnetic resonance in medicine : official journal of the Society of
  Magnetic Resonance in Medicine / Society of Magnetic Resonance in Medicine},
  vol.~61, no.~1, pp. 103--16, Jan. 2009.
  \url{http://www.ncbi.nlm.nih.gov/pubmed/19097216}
\BIBentrySTDinterwordspacing

\bibitem{Jung2010}
\BIBentryALTinterwordspacing
H.~Jung and J.~C. Ye, ``{Motion estimated and compensated compressed sensing
  dynamic magnetic resonance imaging: What we can learn from video compression
  techniques},'' \emph{International Journal of Imaging Systems and
  Technology}, vol.~20, no.~2, pp. 81--98, May 2010.
  \url{http://doi.wiley.com/10.1002/ima.20231}
\BIBentrySTDinterwordspacing

\bibitem{Chen2011}
\BIBentryALTinterwordspacing
L.~Chen, A.~Samsonov, and E.~V.~R. DiBella, ``{A framework for generalized
  reference image reconstruction methods including HYPR-LR, PR-FOCUSS, and k-t
  FOCUSS.}'' \emph{Journal of magnetic resonance imaging : JMRI}, vol.~34,
  no.~2, pp. 403--12, Aug. 2011.
  \url{http://www.pubmedcentral.nih.gov/articlerender.fcgi?artid=3142947\&tool=pmcentrez\&rendertype=abstract}
\BIBentrySTDinterwordspacing

\bibitem{Rauhut2008}
\BIBentryALTinterwordspacing
H.~Rauhut, K.~Schnass, and P.~Vandergheynst, ``{Compressed Sensing and
  Redundant Dictionaries},'' \emph{IEEE Transactions on Information Theory},
  vol.~54, no.~5, pp. 2210--2219, May 2008.
  \url{http://ieeexplore.ieee.org/lpdocs/epic03/wrapper.htm?arnumber=4494699}
\BIBentrySTDinterwordspacing

\bibitem{Candes2010}
E.~J. {Candes}, Y.~C. {Eldar}, D.~{Needell}, and P.~{Randall}, ``{Compressed
  Sensing with Coherent and Redundant Dictionaries},'' \emph{ArXiv e-prints},
  May 2010.

\bibitem{Donoho2009}
\BIBentryALTinterwordspacing
D.~Donoho and J.~Tanner, ``{Observed universality of phase transitions in
  high-dimensional geometry, with implications for modern data analysis and
  signal processing.}'' \emph{Philosophical transactions. Series A,
  Mathematical, physical, and engineering sciences}, vol. 367, no. 1906, pp.
  4273--93, Nov. 2009.  \url{http://www.ncbi.nlm.nih.gov/pubmed/19805445}
\BIBentrySTDinterwordspacing

\bibitem{Boyd2011}
E.~C. B. P. J.~E. Stephen~Boyd, Neal~Parikh, ``Distributed optimization and
  statistical learning via the alternating direction method of multipliers,''
  \emph{Foundations and Trends® in Machine Learning}, 2011.

\bibitem{Goldstein2009}
\BIBentryALTinterwordspacing
T.~Goldstein and S.~Osher, ``The split bregman method for l1-regularized
  problems,'' \emph{SIAM Journal on Imaging Sciences}, vol.~2, no.~2, pp.
  323--343, 2009.  \url{http://link.aip.org/link/?SII/2/323/1}
\BIBentrySTDinterwordspacing

\bibitem{Afonso2011}
\BIBentryALTinterwordspacing
M.~Afonso, J.~Bioucas-Dias, and M.~Figueiredo, ``{An augmented Lagrangian
  approach to the constrained optimization formulation of imaging inverse
  problems},'' \emph{Image Processing, IEEE Transactions on}, no.~99, pp. 1--1,
  2011.  \url{http://ieeexplore.ieee.org/xpls/abs\_all.jsp?arnumber=5570998}
\BIBentrySTDinterwordspacing

\bibitem{Bioucas2007}
J.~Bioucas-Dias and M.~Figueiredo, ``A new twist: Two-step iterative
  shrinkage/thresholding algorithms for image restoration,'' \emph{Image
  Processing, IEEE Transactions on}, vol.~16, no.~12, pp. 2992 --3004, dec.
  2007.

\bibitem{Beck2009}
\BIBentryALTinterwordspacing
A.~Beck and M.~Teboulle, ``{Fast gradient-based algorithms for constrained
  total variation image denoising and deblurring problems.}'' \emph{IEEE
  transactions on image processing : a publication of the IEEE Signal
  Processing Society}, vol.~18, no.~11, pp. 2419--34, Nov. 2009.
  \url{http://www.ncbi.nlm.nih.gov/pubmed/19635705}
\BIBentrySTDinterwordspacing

\bibitem{Afonso2010a}
\BIBentryALTinterwordspacing
M.~Afonso, J.~Bioucas-Dias, and M.~Figueiredo, ``{Fast image recovery using
  variable splitting and constrained optimization},'' \emph{Image Processing,
  IEEE Transactions on}, vol.~19, no.~9, pp. 2345--2356, Sep. 2010.
  \url{http://www.ncbi.nlm.nih.gov/pubmed/20378469
  http://ieeexplore.ieee.org/xpls/abs\_all.jsp?arnumber=5445028}
\BIBentrySTDinterwordspacing

\bibitem{Akcakaya2010}
\BIBentryALTinterwordspacing
M.~Akcakaya and V.~Tarokh, ``{Shannon-Theoretic Limits on Noisy Compressive
  Sampling},'' \emph{IEEE Transactions on Information Theory}, vol.~56, no.~1,
  pp. 492--504, Jan. 2010.
  \url{http://ieeexplore.ieee.org/lpdocs/epic03/wrapper.htm?arnumber=5361481}
\BIBentrySTDinterwordspacing

\bibitem{Ramani2011}
\BIBentryALTinterwordspacing
S.~Ramani and J.~Fessler, ``{Parallel MR Image Reconstruction Using Augmented
  Lagrangian Methods},'' \emph{Medical Imaging, IEEE Transactions on}, vol.~30,
  no.~3, pp. 694--706, 2011.
  \url{http://ieeexplore.ieee.org/xpls/abs\_all.jsp?arnumber=5639083}
\BIBentrySTDinterwordspacing

\bibitem{Thirion1996}
\BIBentryALTinterwordspacing
J.~Thirion, ``{Non-rigid matching using demons},'' in \emph{cvpr}.\hskip 1em
  plus 0.5em minus 0.4em\relax Published by the IEEE Computer Society, 1996, p.
  245.
  \url{http://www.computer.org/portal/web/csdl/doi/10.1109/CVPR.1996.517081}
\BIBentrySTDinterwordspacing

\bibitem{Pennec1999}
\BIBentryALTinterwordspacing
X.~Pennec, P.~Cachier, and N.~Ayache, ``{Understanding the “Demon’s
  Algorithm”: 3D Non-rigid Registration by Gradient Descent},'' in
  \emph{Medical Image Computing and Computer-Assisted Intervention --
  MICCAI’99}, vol. 1679.\hskip 1em plus 0.5em minus 0.4em\relax Springer
  Berlin / Heidelberg, 1999, pp. 597--605.
  \url{http://www.springerlink.com/index/28v36g511w845142.pdf
  http://dx.doi.org/10.1007/10704282\_64}
\BIBentrySTDinterwordspacing

\bibitem{Vercauteren2009}
\BIBentryALTinterwordspacing
T.~Vercauteren, X.~Pennec, A.~Perchant, and N.~Ayache, ``{Diffeomorphic demons:
  efficient non-parametric image registration.}'' \emph{NeuroImage}, vol.~45,
  no. 1 Suppl, pp. S61--72, Mar. 2009.
  \url{http://www.ncbi.nlm.nih.gov/pubmed/19041946}
\BIBentrySTDinterwordspacing

\bibitem{Gee1994}
\BIBentryALTinterwordspacing
J.~Gee, D.~Haynor, M.~Reivich, R.~Bajcsy, and Others, ``{Finite element
  approach to warping of brain images},'' \emph{Medical Imaging}, pp. 327--337,
  1994.
  \url{http://citeseerx.ist.psu.edu/viewdoc/download?doi=10.1.1.45.5852\&amp;rep=rep1\&amp;type=ps}
\BIBentrySTDinterwordspacing

\bibitem{Samavati2009}
\BIBentryALTinterwordspacing
N.~Samavati, ``{Deformable Multi-Modality Image Registration Based on Finite
  Element Modeling and Moving Least Squares},'' Ph.D. dissertation, 2009.
  \url{http://digitalcommons.mcmaster.ca/opendissertations/4371/}
\BIBentrySTDinterwordspacing

\end{thebibliography}

\end{document}